\documentclass[fleqn,usenatbib]{mnras}

\usepackage{newtxtext,newtxmath}

\usepackage[T1]{fontenc}
\usepackage{ae,aecompl}
\usepackage{appendix}

\usepackage{graphicx}	
\usepackage{amsmath}	
\usepackage{amssymb}	
\usepackage{subfig}
\usepackage{color}
\usepackage[utf8]{inputenc}
\usepackage[french,german,spanish,english]{babel}
\usepackage[normalem]{ulem}

\newcommand{\msol}{\rm M_{\odot}}
\newcommand{\mh}{\rm M_{\rm halo}}
\newcommand{\rtwo}{\rm r_{\rm 200}}

\newcommand{\fesc}{\rm f_{\rm esc}}
\newcommand{\fescr}{\rm f^{\rm ray}_{\rm esc}}

\newcommand{\fescsub}{\rm f^{\rm sub}_{\rm esc}}
\newcommand{\fescnet}{\rm f^{\rm net}_{\rm esc}}

\newcommand{\xhi}{\rm X_{\rm HI}}

\newcommand{\Lesc}{$\rm L_{\rm esc}$ }
\newcommand{\Lescine}{{\rm L_{\rm esc}} }

\newcommand{\magn}{$\rm M_{\rm AB1600}$ }

\title[CoDa II galactic photon budget]{Galactic ionising photon budget during the Epoch of Reionisation in the Cosmic Dawn II simulation.}

\author[Joseph S. W. Lewis]{
Joseph S. W. Lewis$^{1}$,
Pierre Ocvirk$^{1}$,
Dominique Aubert$^{1}$,
Jenny G. Sorce$^{2,3,4}$,
\newauthor
Paul R. Shapiro$^{5}$,
Nicolas Deparis$^{1}$,
Taha Dawoodbhoy$^{5}$,
Romain Teyssier$^{6}$,
\newauthor
Gustavo Yepes$^{7,8}$,
Stefan Gottlöber$^{4}$,
Kyungjin Ahn$^{9}$,
Ilian T. Iliev$^{10}$,
Jonathan Chardin$^{1}$
\\
$^{1}$Observatoire Astronomique de Strasbourg, Université de Strasbourg, CNRS UMR 7550, 11 rue de l’Université, 67000 Strasbourg, France\\
$^{2}$Univ. Lyon, ENS de Lyon, Univ. Lyon I, CNRS, Centre de Recherche Astrophysique de Lyon, UMR5574, F-69007, Lyon, France \\
$^{3}$Univ. Lyon, Univ. Lyon I, ENS de Lyon, CNRS, Centre de Recherche Astrophysique de Lyon, UMR5574, F-69230, Saint-Genis-Laval, France\\
$^{4}$Leibniz-Institut für Astrophysik Potsdam (AIP), An der Sternwarte 16, D-14482 Potsdam, Germany\\
$^{5}$Department of Astronomy, University Texas, Austin, TX 78712-1083, USA\\
$^{6}$Institute for Theoretical Physics, University of Zurich, Winterthurerstrasse 190, CH-8057 Zürich, Switzerland\\
$^{7}$ Departamento de F\'isica Te\'orica M-8, Universidad Aut\'onoma de Madrid, Cantoblanco, 28049, Madrid, Spain\\
$^{8}$ Centro de Investigaci\'on Avanzada en F\'isica  Fundamental (CIAFF), Universidad Aut\'onoma de Madrid, 28049 Madrid, Spain\\
$^{9}$Chosun University, 375 Seosuk-dong, Dong-gu, Gwangjiu 501-759, Korea\\
$^{10}$Astronomy Center, Department of Physics \& Astronomy, Pevensey II Building, University of Sussex, Falmer, Brighton BN1 9QH, United Kingdom
}
\date{Accepted XXX. Received YYY; in original form ZZZ}

\pubyear{2019}

\begin{document}
\label{firstpage}
\pagerange{\pageref{firstpage}--\pageref{lastpage}}
\maketitle

\begin{abstract}
  Cosmic Dawn (“CoDa”) II yields the first statistically-meaningful determination of the relative contribution to reionization by galaxies of different halo mass, from a fully-coupled radiation-hydrodynamics simulation of the epoch of reionization large enough ($\sim 100$ Mpc) to model global reionization while resolving the formation of all galactic halos above  $\sim 10^8 \msol$. Cell transmission inside haloes is bi-modal -- ionised cells are transparent, while neutral cells absorb the photons their stars produce - and the halo escape fraction $ \fesc$ reflects the balance of star formation rate ("SFR") between these modes. The latter is increasingly prevalent at higher halo mass, driving down $\fesc$ (we provide analytical fits to our results), whereas halo escape luminosity, proportional to $\fesc \times {\rm SFR}$, increases with mass. Haloes with dark matter masses within $6\times 10^{8} \msol < \mh < 3 \times 10^{10} \msol$ produce $\sim 80$\% of the escaping photons at z=7, when the Universe is 50\% ionised, making them the main drivers of cosmic reionization. Less massive haloes, though more numerous, have low SFRs and contribute less than 10\% of the photon budget then, despite their high $\fesc$. High mass haloes are too few and too opaque, contributing $<10$\% despite their high SFRs. The dominant mass range is lower (higher) at higher (lower) redshift, as mass function and reionization advance together (e.g. at z$=8.5$, x$_{\rm HI}=0.9$, $\mh < 5\times 10^9 \msol$ haloes contributed $\sim 80$\%). Galaxies with UV magnitudes \magn between $-12$ and $-19$ dominated reionization between z$=6$ and 8.
  
\end{abstract}




\section{Introduction}

Current observations are consistent with the hypothesis that the Universe was reionized when UV starlight from massive stars escaped from the early galaxies in which they formed, creating intergalactic  H II regions that grew in size and number until they overlapped to fully-ionise the intergalactic medium (hereafter IGM) by z $=$ 6. [For reviews and references, see, for instance, \cite{dayal_early_2018} and \cite{barkana_physics_2007}]. During the epoch of reionization (“EOR”), the globally-averaged ionised fraction was equivalent to the volume filling factor of these H II regions, which increased monotonically in an evolving patchwork of fully-ionised and fully-neutral zones.  How fast this volume filling factor grew was determined primarily by the average balance between the rate of release of ionising photons by galaxies and the recombination rate of H atoms in their surrounding IGM.   The release rate in a given patch depended upon the galaxy formation rate there, the star formation rates (hereafter “SFR”) inside each galaxy, the spectra and luminosities of those stars, and the galactic escape fractions $\fesc$ of their ionising photons.  The recombination rate in the surrounding IGM depended upon its evolving inhomogeneous density field.   All these ingredients varied in space and time in a complex way. Firstly, since structure formation was inhomogeneous. Secondly, because Reionization and the energy release associated with the star formation that drives it exerted hydrodynamical feedback; therefore, the ingredients were inter-dependent. 

To predict their coupled evolution in the context of $\Lambda$CDM cosmology, to test the latter against observations, we must model the gravitational and gas dynamics of dark and baryonic matter and the radiative transfer of ionising radiation in and between galaxies as they form. To capture the large-scale structure of inhomogeneous reionization, this must be done in a representative volume large enough ($\sim$ 100 Mpc), and with enough resolving power to form all the galaxies in that volume which contribute to reionization.    As the dominant contributors are thought to be the “atomic-cooling haloes” (hereafter “ACHs”) -- those of virial temperatures above $10^4$ K and masses above $10^8 \msol$ -- this means we must be able to resolve the formation of all the millions of haloes above $10^8 \msol$ in that large volume\footnote{Although the first stars form in mini-haloes (hereafter “MHs”) (i.e. those with halo mass M$< 10^8 \msol$ and virial temperatures T$<10^4$K) that can cool gas by H$_2$ molecular cooling \citep[some stars may even form in metal-cooling MHs, as shown in][]{wise_birth_2014}, the rising UV background during the EOR limits their contribution to the earliest stages of reionization.  As a result, their relative contribution when compared to the more massive ACHs appears small ( $\sim$ 10 - 20\%), \citep{kimm_feedback-regulated_2017,ahn_detecting_2012}.}.
In principle, to capture the full details of star formation within each galaxy, we would also have to resolve the interstellar medium of each galaxy down to the sub-parsec scale on which molecular clouds fragment into protostars which then collapse into stars. The latter is currently out-of-reach computationally, however, even in the highest-resolution simulations to-date of a single galaxy. This means star formation and its local energy release must generally be treated as a “sub-grid” process.

We have developed the Cosmic Dawn (hereafter “CoDa”) Project, to simulate reionization and galaxy formation together, self-consistently, with fully-coupled, radiation-hydrodynamics, on a large-enough scale and with sufficient mass resolution to satisfy these requirements \citep{ocvirk_cosmic_2016,aubert_inhomogeneous_2018,dawoodbhoy_suppression_2018,ocvirk_cosmic_2018}. CoDa I (91 Mpc box), described in \cite{ocvirk_cosmic_2016} and \cite{dawoodbhoy_suppression_2018}, and CoDa II (94.5 Mpc box), described in \cite{ocvirk_cosmic_2018}, both used the massively-parallel, hybrid CPU-GPU code RAMSES-CUDATON on a uniform grid of $4096^3$ cells for the baryons and the radiation field, with $4096^3$ N-body particles for the dark matter.  CoDa I-AMR  (91 Mpc), on the other hand, used another massively-parallel, hybrid CPU-GPU code EMMA \citep{aubert_emma:_2015}, with Adaptive Mesh Refinement (“AMR”), with $2048^3$ particles on a grid of $2048^3$ coarse cells from which AMR increased the resolution locally, by up to a factor of 8, to follow the increasing local overdensity, leading to 18 billion cells after refinement.   All three CoDa simulations were in volumes large enough to model the inhomogeneity and globally-averaged time-history of reionization, while also serving to model reionization and galaxy formation in the Local Universe, by adopting “constrained realisations” of the Gaussian random noise initial conditions which were derived from galaxy survey data so as to reproduce the familiar structures of the Local Universe, such as the MW, M31, and the Virgo cluster, when evolved forward to z$=0$ \citep{sorce_cosmicflows_2016}.   We refer the reader to the papers cited above to describe our CoDa simulations and their relative differences in more detail.   Our purpose here is to use the most recent of them, CoDa II, to find the ionising luminosities of all the galaxies that formed in it during the EOR, to make the first statistically-meaningful determination of the relative contribution to reionization by galaxies of different halo mass, over the full range of masses that contribute significantly, in a fully coupled radiation hydro-dynamical numerical simulation.

The escape fraction $\fesc$ of galaxies is difficult to observe directly. Indeed, the individual galaxies must be bright enough to detect, but also to compare their fluxes and spectral information at different wavelengths (above and below the H Lyman-limit). This must then be interpreted in terms of a model in which stars are assumed to have some initial mass function (hereafter “IMF”) and a SFR, which determines their spectral energy distribution (hereafter “SED”) over time.   The radiation that emerges from the galaxy at wavelengths longward of the Lyman limit  is then assumed to be a combination of this SED and the nebular emission which results from re-processing the absorbed fraction of ionising starlight by the interstellar gas, and may be partially attenuated by internal dust.  Starlight emitted blueward of the H Lyman limit is attenuated by photoionizing H atoms in the ISM of the galaxy and possibly attenuated further by dust. These processes are reflected in the net absorbed fraction ($1 - \fesc$), which may also include attenuation by the bound-free opacity of foreground Lyman limit absorbers along the line of sight.    Observations of galaxies at different redshifts face different challenges, as they involve different spectral regions depending on z. Moreover, the foreground opacity is also a strong function of increasing redshift. As a consequence, observational determinations of $\fesc$ are few and still uncertain. A review of this subject is well beyond the scope of this paper; the reader is referred to, e.g., \cite{izotov_eight_2016}, and the review by \cite{dayal_early_2018} and references therein for a summary.  

{ On the theory side, results are also rather limited.   Some attempt to derive an empirical $\fesc$, one-size-fits-all. For an assumed form and amplitude of the UV luminosity function of galaxies above redshift 6, they proceed by adjusting $\fesc$ to release enough ionising photons to finish reionization in time to satisfy various observational constraints (We refer to these efforts as “one-zone” models e.g. \citet{robertson_cosmic_2015}).    In doing so values like $\fesc=$ 10 or 20\% are sometimes reported, but this depends strongly on the underlying assumptions that led to it. These models often requires some redshift evolution \citep[for instance, in][]{puchwein_consistent_2019,haardt_high-redshift_2015}.  Other attempts to determine a global value for $\fesc$ employ semi-analytical or semi-analytical models of reionization. Again, these are similarly adjusted to match observational constraints, but use the model’s own statistical determinations of the rate of formation of galactic halos from cosmological initial conditions, and some assumption about the SFR per halo \citep[e.g.][]{dayal_reionization_2020,ferrara_escape_2013}.   A variety of galaxy formation simulations also exist which attempt to predict the $\fesc$ and SFR from their simulated galaxies. However, these are mostly without radiative transfer or only post-processed with radiative transfer \citep[Such as in][]{yajima_escape_2011,razoumov_ionizing_2010,paardekooper_first_2015,ma_difficulty_2015,anderson_little_2017}. 
Recent studies do account for fully-coupled radiation-hydrodynamics, with a focus on spatial resolution, considering a single galaxy or a fraction thereof \citep{trebitsch_fluctuating_2017,trebitsch_escape_2018,kimm_escape_2014,kimm_understanding_2019,trebitsch_obelisk_2020,yoo_origin_2020}. This focus on resolving internal galactic structure precludes the follow-up of ionizing radiation propagation at cosmological scales.}

A review of this subject, too,  is well beyond the scope of this paper; the reader is again referred to \cite{dayal_early_2018} for a summary and references, but we will describe some of these results in what follows, as we compare with our own.    

The relative contribution of different mass halos to the total ionizing photon budget released into the IGM during the EOR depends, not only upon the values of $\fesc$ for each galaxy, but on their SFRs and and the evolution of their population, as well. These aspects combine to determine the ionizing luminosity function of galaxies.   In this work, we shall investigate this galactic ionizing photon budget.  Since star formation efficiency typically rises with halo mass within the range of masses we represent \citep[not only in CoDa II, but as generally expected, see][]{moster_galactic_2013,legrand_cosmos-ultravista_2019}, while the abundance of haloes decreases with halo mass \citep[e.g. see the halo mass functions of][]{sheth_ellipsoidal_2001,watson_halo_2013}, the mass range of contributing haloes may, in principle, be broad, with a maximum contribution from halos that, at different redshifts, may be anywhere within the broad range $10^8 \msol < {\rm M_{halo}} < 10^{12} \msol$.

Previous work on the role of simulated galaxies in ionising the IGM has sometimes been difficult to reconcile. However, it seems that with the recent advent of higher resolution and of fully-coupled simulations, a few elements of consensus have begun to emerge.  Studies such as \cite{anderson_little_2017} and \cite{yajima_escape_2011} find that reionization is driven by the more numerous, low-mass galaxies  (${\rm M_{halo}} < 10^{9.5} \msol$), which broadly agrees with the conclusions of \cite{kimm_escape_2014},  who find that the photon budget is dominated by masses $10^8 \msol < {\rm M_{halo}} < 10^9 \msol$ before z=8, after which more massive haloes take over. Similarly, a more recent effort by \cite{katz_census_2018} seems to favour haloes within the range $10^9 \msol < {\rm M_{halo}} < 10^{10} \msol$ during the EOR, and those of higher mass at z = 6.

Most of the previous simulations are in volumes which are not large enough (most are boxes smaller than 25 Mpc across) to fully represent the halo mass function above $\sim 10^{10} \msol$.   They may therefore be missing some of the largest haloes and galaxies, the ones, in fact, that form the most stars. The contributions of these highest-mass haloes ($>10^{10} \msol$) to the photon budget in these studies is, therefore, partially absent. This could have a further, profound effect, on the rate LyC photons are released from the lowest-mass galaxies, the ones that are reionized and suppressed by external sources, as well as dramatically alter the geometry of ionised regions throughout the EOR.  Moreover, when reionization is simulated in too small a box, the duration of reionization is too small compared with that found in a volume large enough to capture the globally-averaged history. In turn, this can affect the relative importance of halos of different mass as their relative abundances evolve with redshift \citep{iliev_simulating_2006,iliev_simulating_2014}, alongside their importance to Reionization.

To overcome these limitations, we will address the photon budget of galaxies during the EOR using the CoDa II simulation \citep{ocvirk_cosmic_2018}, produced with the RAMSES-CUDATON code \citep{ocvirk_cosmic_2016}, which couples RAMSES \citep{teyssier_cosmological_2002}, the code for baryonic hydrodynamics and dark matter N-body dynamics, to ATON \citep{aubert_radiative_2008}, the code for radiative transfer of ionising radiation and non-equilibrium ionisation rate equations.   The CoDa II simulation ran from z=150 to z=5.8 in a comoving cubic box 94.533 Mpc on a side, with a high-enough mass resolution to form every galaxy in that volume with halo mass above $10^8 \msol$. This is sufficient to satisfy the requirement for large enough volume to simulate the EOR and its inhomogeneity, with a statistically-meaningful halo mass function over the full mass range that may contribute significantly to reionization. Further featuring the fully-coupled radiation-hydrodynamics (including radiation transport at the full speed of light) necessary to study the release of ionising starlight into the IGM by galaxies during the EOR and its transport between galaxies involving highly-supersonic ionisation fronts. This combination of very large volume with complete sampling of the galactic sources within it represents a necessary compromise.  The focus on large scales comes at a cost; we do not attempt to achieve the higher resolution inside galaxies that some other recent simulations do [e.g.\citep{trebitsch_fluctuating_2017,rosdahl_sphinx_2018,kimm_understanding_2019,katz_census_2018,katz_tracing_2019,trebitsch_obelisk_2020}], which may affect some of the internal halo physics that are important for our problem, such as the escape of ionising photons. One of the goals of this study, however, is to demonstrate that, despite our relatively coarser spatial resolution internal to individual galaxies in CoDa II , the "global" halo quantities relevant for describing the radiative properties of high-redshift galaxies, such as their escape fraction and total escape luminosity in ionizing photons, are meaningful and well-captured, thereby validating the CoDa II -like approach, and paving the way towards even larger numerical simulations of the EOR with it in the future.

In this paper, we first, in Sect. \ref{sec:methods}, outline our numerical approach and computations. Then, in Sect. \ref{sec:escape}, we lay out our escape fraction results. Then in Sec. \ref{sec:budget}, we present the ionising galactic photon budget. Finally, in Sect. \ref{sec:conclusion}, we summarise our findings and propose some directions in which to take our subsequent efforts.

\section{Methods}

\label{sec:methods}

\subsection{Cosmic Dawn Simulations}

CoDa I\ and CoDa II \ are the largest coupled radiation hydrodynamics cosmological grid-based simulations aimed at studying the EoR. In the simulation code RAMSES-CUDATON, the RAMSES hydrodynamics+N-body code \citep{teyssier_cosmological_2002} and the ATON radiative transfer code \citep{levermore_relating_1984,aubert_radiative_2008} are coupled, forming a hybrid code : hydrodynamics, gravitation, star formation, and supernova feedback are managed by the central processing units (CPUs), while the more computationally intensive mono-group\footnote{Effective photon energy 20.28 eV} radiative transfer, Hydrogen photo-chemistry and cooling are managed by the graphics processing units (GPUs).
The resulting acceleration allows us to perform simulations using the full speed of light, thereby circumventing possible artefacts arising from the use of a reduced speed of light (see \citet{gnedin_proper_2016,deparis_impact_2019,ocvirk_impact_2019,wu_simulating_2019} for details on the impact of reduced or variable speed of light approximations). 

{ We do not account for chemical enrichment, nor dust. The simulation focuses on the stellar component's ability to drive cosmic reionization, and therefore active galactic nuclei formation, feedback and ionizing emissivity are not taken into account.}

{ The large box size (94.44 cMpc box) and relatively high resolution for a simulation of cosmic reionization ($4096^3$ dark matter particles and $4096^3$ cells) yield a comoving cell size of 23.06 ckpc (i.e. 3.3 kpc physical at z=6) and a dark matter (stellar) particle mass of $4.07 \times 10^{5}  \msol$ (11732 $\msol$). } With these specifications, CoDa II can represent the large-scale spatial variance in the reionisation process whilst self-consistently dealing with the physics of the haloes that interest us (ie : those within $10^8 \msol \lesssim \mh \lesssim 10^{12} \msol$), and providing us a huge sample of galactic haloes: there are around 13 million dark matter halos at the end of the EoR in CoDa II.

CoDa II is compatible with a number of observational constraints related to the EoR, most notably the reionization history of the Universe, the Thomson optical depth measured from the cosmic microwave background, and the UV luminosity function of galaxies, as shown in \citet{ocvirk_cosmic_2018}. For further information relating to the code's design, setup, and runs, we refer the reader to \citet{ocvirk_cosmic_2018}.

 \subsection{Halo detection and boundaries}

\label{s:nint}
Dark matter haloes (haloes throughout the text) are detected using a Friends-of-Friends algorithm described in \citet{roy_pfof:_2014}, which produces a catalogue of haloes with their positions and masses $\mh$. We can define a halo's virial radius, based on it's mass $\mh$, as $\rtwo$, as in \citet{ocvirk_cosmic_2016,ocvirk_cosmic_2018}:

   \begin{equation}  
      \mh = \frac{4}{3} \pi \, r_{200}^3\times 200 \, \bar  {\rm \rho}_{\rm DM}\, ,
      \label{eq:2}
    \end{equation}
where $\bar{\rm \rho}_{\rm DM}$ is the average cosmic dark matter density.

{ As in \citet{ocvirk_cosmic_2016,ocvirk_cosmic_2018}, we assume that one galaxy resides in each dark matter halo, and that the $\rtwo$ limit of the halo is the limit of the galaxy. This assumption is valid in the vast majority of cases, in which the star forming region of each halo has one clear stellar mass peak within $\rtwo$.}

\subsection{Halo escape fraction: ray-tracing, sub-grid and net}

Using CoDa II's gas density and ionisation fields, we can compute the optical depths encountered by photons emitted in the halo along paths, from their injection to $\rtwo$. For a given halo, and for a given halo cell, we use the $healpix$ python module $healpy$ to sample the $\rtwo$ sphere with 768 evenly distributed end points (We pick this number so as to adequately resolve our largest haloes. Increasing the number of rays by a factor of two only yields a difference of the order of $10^{-4}$ in $\fescr$ for the most error susceptible computation). We then compute the optical depth along the path from the source cell centre to each end point as :
\begin{equation}
    \rm \tau_{\rm path}= \int_{path} \sigma_E\times \rho_{\rm HI}\times {\rm dl} \, ,
\end{equation}  
where $\sigma_E=2.493 \times 10^{-22} {\rm m}^2$ is the effective Hydrogen ionisation cross-section of the photon group considered here in CoDa II, $\rho_{\rm HI}$ is the neutral Hydrogen physical density of the cell, and dl is an element of length.

The fraction of photons reaching $\rtwo$ from a cell, $\rm Tr_{200}$, is then obtained as the average of the transmissions along all 768 paths connecting the cell to the $\rtwo$ sphere: $\rm Tr_{200}=\,\langle\exp{(-\tau_{\rm path})}\rangle_{\rm paths}$ where the bracket denotes the average over all 768 paths.
Fig. \ref{fig:fesc_draw} is provided as an explanatory figure in Appendix \ref{sec:fesc_draw}.

Finally, the halo escape fraction $\fescr$ is obtained as the SFR-weighted average of the transmissions of all the emitting cells of the halo, i.e. for a halo containing N cells indexed by the integer ${\rm i}$:
\begin{equation}
    \fescr = \frac{\sum_{\rm i}{\rm SFR_i 
    Tr_{200,i}}}{{\rm \sum_i SFR_i}} .
\end{equation}

 The star formation rate SFR of a cell is obtained as the stellar mass formed in the last 10 Myr, ${\rm M_{stars}(<10 Myr)}$ divided by a 10 Myr duration:
 \begin{equation}
     {\rm SFR = M_{\rm stars} (<10 Myr) / 10^7} \, \msol/yr ,
     \label{eq:SFR}
 \end{equation}
 In the rest of the paper, "star-forming" (halo or cell) always means that some stellar mass has been formed in the last 10 Myr, i.e. ${\rm M_{\rm stars} (<10 Myr) }>0$.
 The "ray" superscript is used to clarify at all times that $\fescr$ is obtained via ray-tracing. We will refrain from discussing the ISM (Inter-Stellar Medium) or CGM nature of the absorbing material. Since these are difficult to separate in CoDa II, we prefer the more general term "halo escape fraction" for $\fescr$. 
 
 Furthermore, since CoDa II does not resolve the ISM of our galaxies, CoDa II also uses a sub-grid escape fraction $\fescsub=0.42$ (chosen in order to obtain an EoR ending around z=6), which accounts for the photons lost to the star's birth cloud and the ISM, i.e. only 0.42 of the ionising photons produced by a stellar particle is deposited in the cell containing it. We can now also define the net halo escape fraction $\fescnet$ as:
 \begin{equation}
     \fescnet=\fescsub\times \fescr \, ,
     \label{eq:fesc_def}
 \end{equation}
 This is the fraction of a halo's stellar population photon production which manages to reach $\rtwo$. Since $\fescsub$ is constant by construction in CoDa II, we focus mostly in the rest of the paper on the determination and behaviour of $\fescr$.

\subsection{Escape luminosity}

To obtain the instantaneous amount of produced ionising photons within a given halo $\dot{\rm \eta}_{\star}$, we sum the ionising photon production of all emitting star particles (i.e. younger than 10 Myr) within $\rtwo$, i.e. 
\begin{equation}
    \dot{\rm \eta}_{\star} = {\rm M_{\rm stars} (<10 Myr)} \times  4.32 \times 10^{46} \, {\rm ph/s} \, , 
\end{equation}
where the factor $4.32 \times 10^{46}$ is the stellar ionising emissivity in ph/s/$\msol$ \footnote{this value is computed from the number of ionising photons per stellar baryon produced by a binary stellar population of metallicity Z=0.001 using the BPASS \citep{eldridge_binary_2017} models, and with a Kroupa IMF \citep{kroupa_variation_2001}} \citep[taken from Tab. 1 of][]{ocvirk_cosmic_2018}. We then define the escape luminosity (\Lesc) of a halo as the product of the intrinsic luminosity and the net halo escape fraction. This gives the number of ionising photons that reach $\rtwo$ per second.
\begin{equation}
    \Lescine = \dot{\rm \eta}_{\star}\times \fescsub\times \fescr \, ph/s,
    \label{eq:Lesc}
\end{equation}
Note that, with these definitions, \Lesc is, as expected, proportional to SFR and $\fescr$.
If we interpret $\rtwo$ as the limit between the halo and the IGM, \Lesc is effectively an estimate of the halo luminosity exiting the halo and entering the IGM. Similarly, we can define the {\rm cell} escape luminosity, by considering $\dot{\rm \eta}_{\star}$ as the photon production within that cell and the average transmission ${\rm Tr_{200}}$ of the paths from that cell to the $\rtwo$ sphere of its host halo.

\section{Results: Halo escape fractions}
\label{sec:escape}
\subsection{Halo escape fraction as a function of mass and redshift}

\begin{figure*}

  \includegraphics[width=0.9\columnwidth]{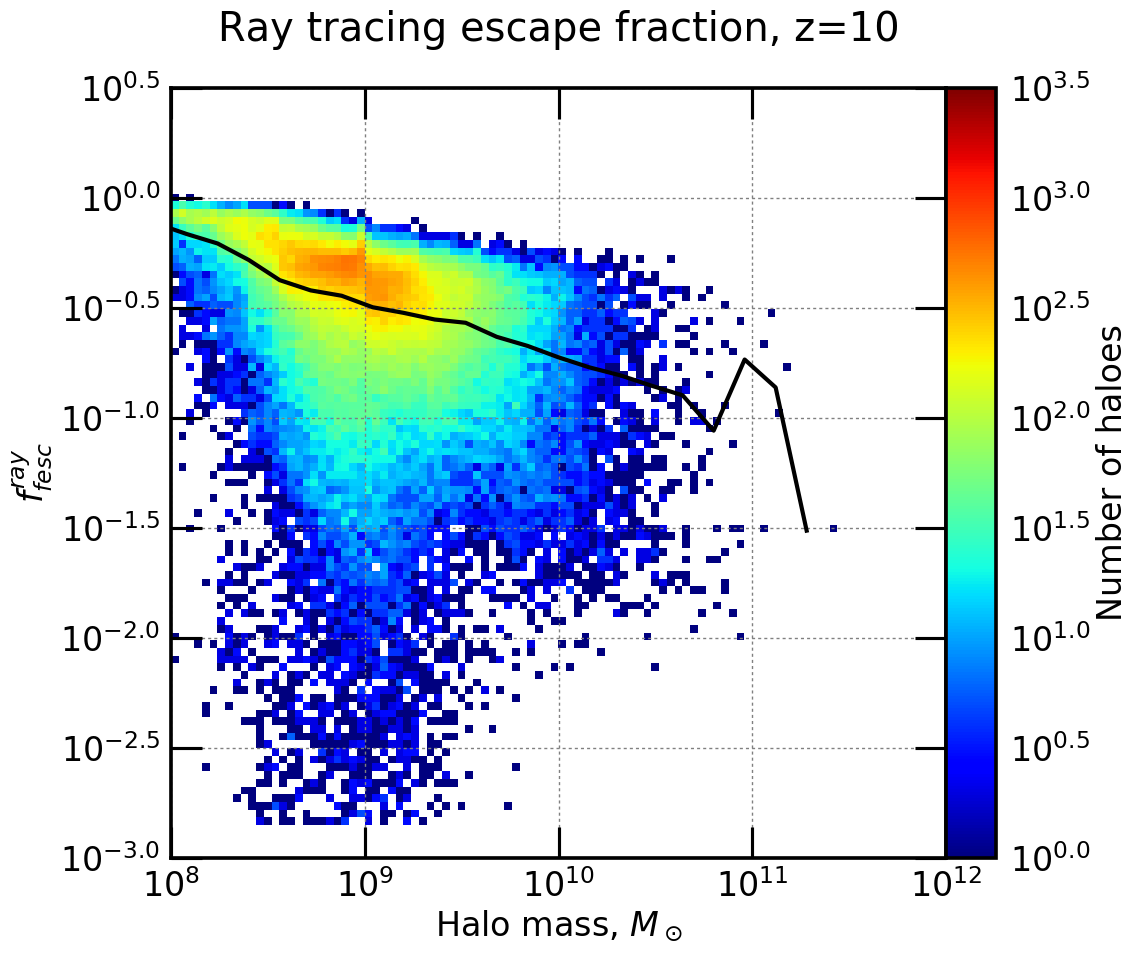}
  \includegraphics[width=0.9\columnwidth]{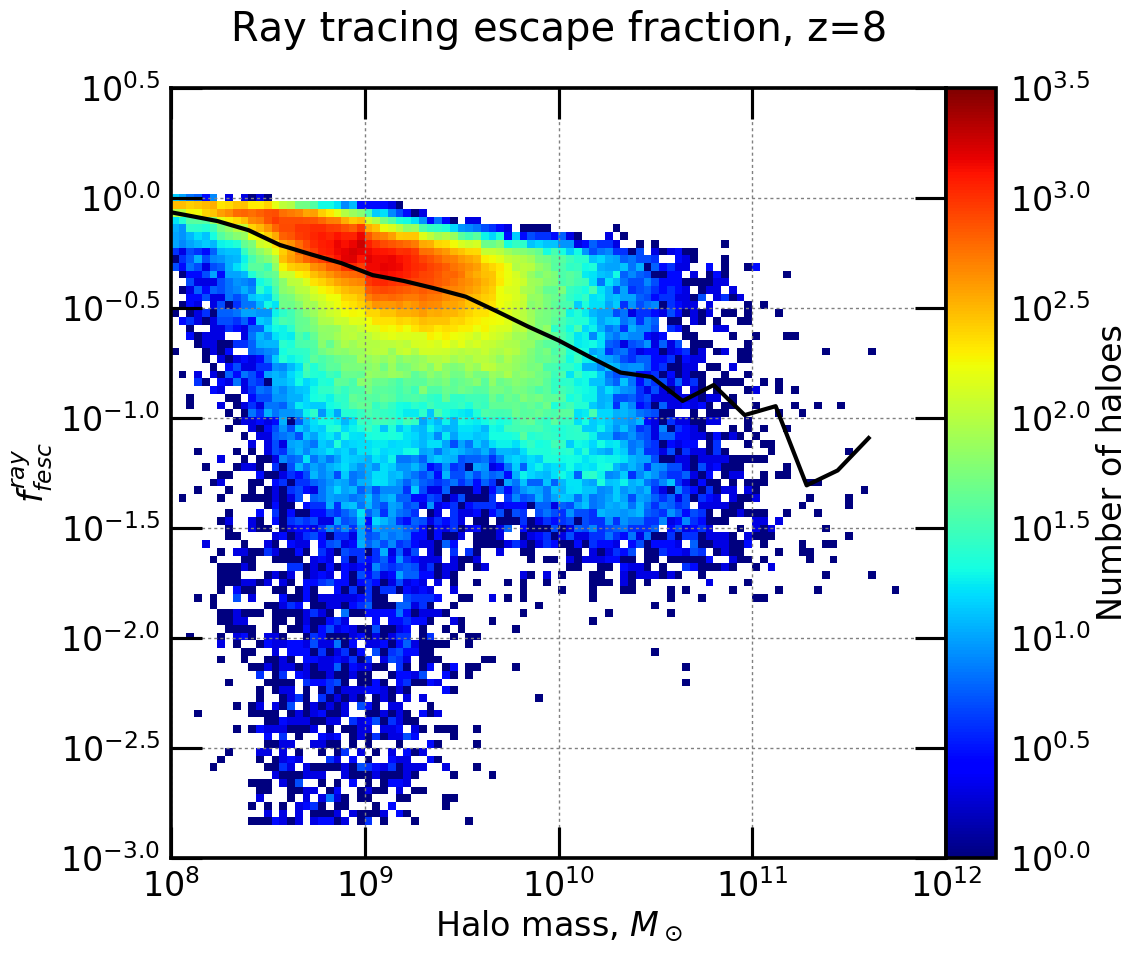}
  \includegraphics[width=0.9\columnwidth]{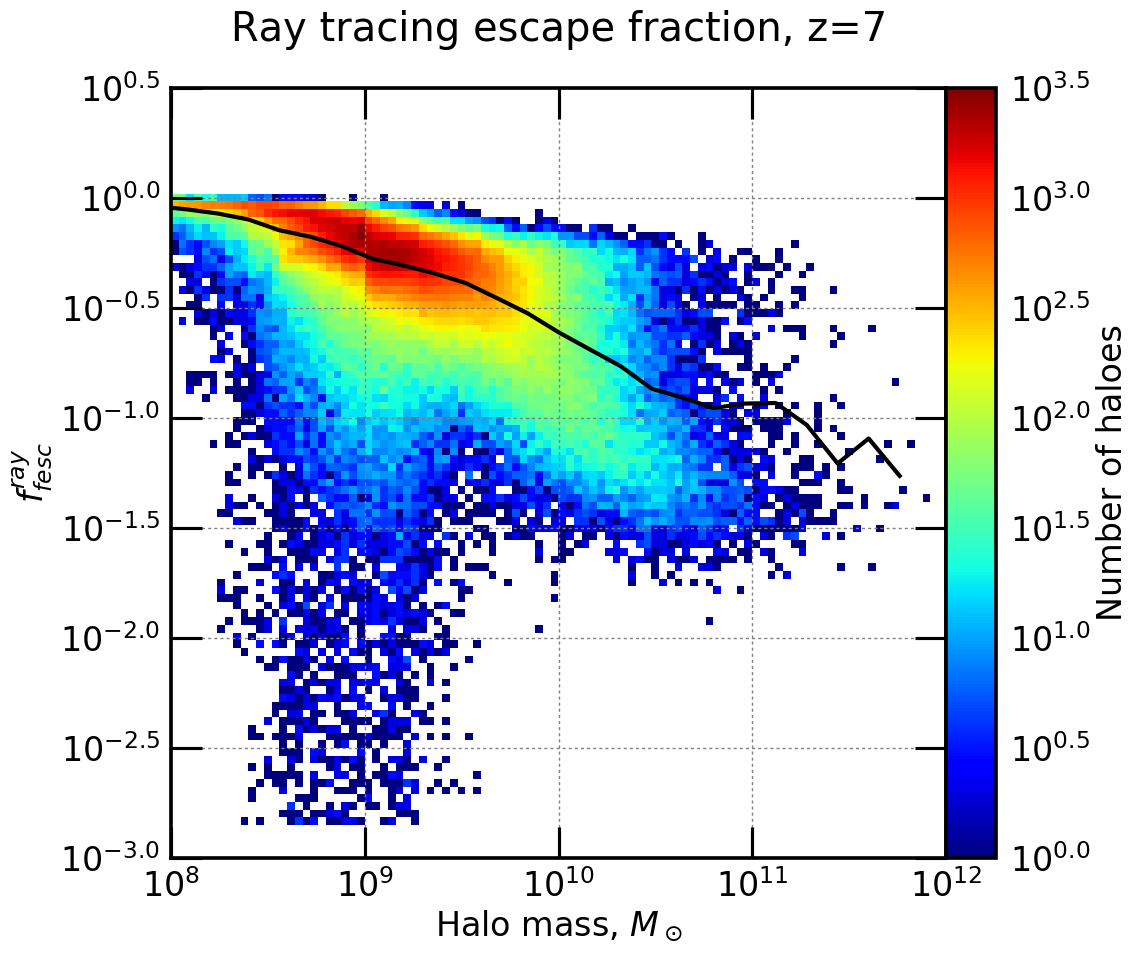}
  \includegraphics[width=0.9\columnwidth]{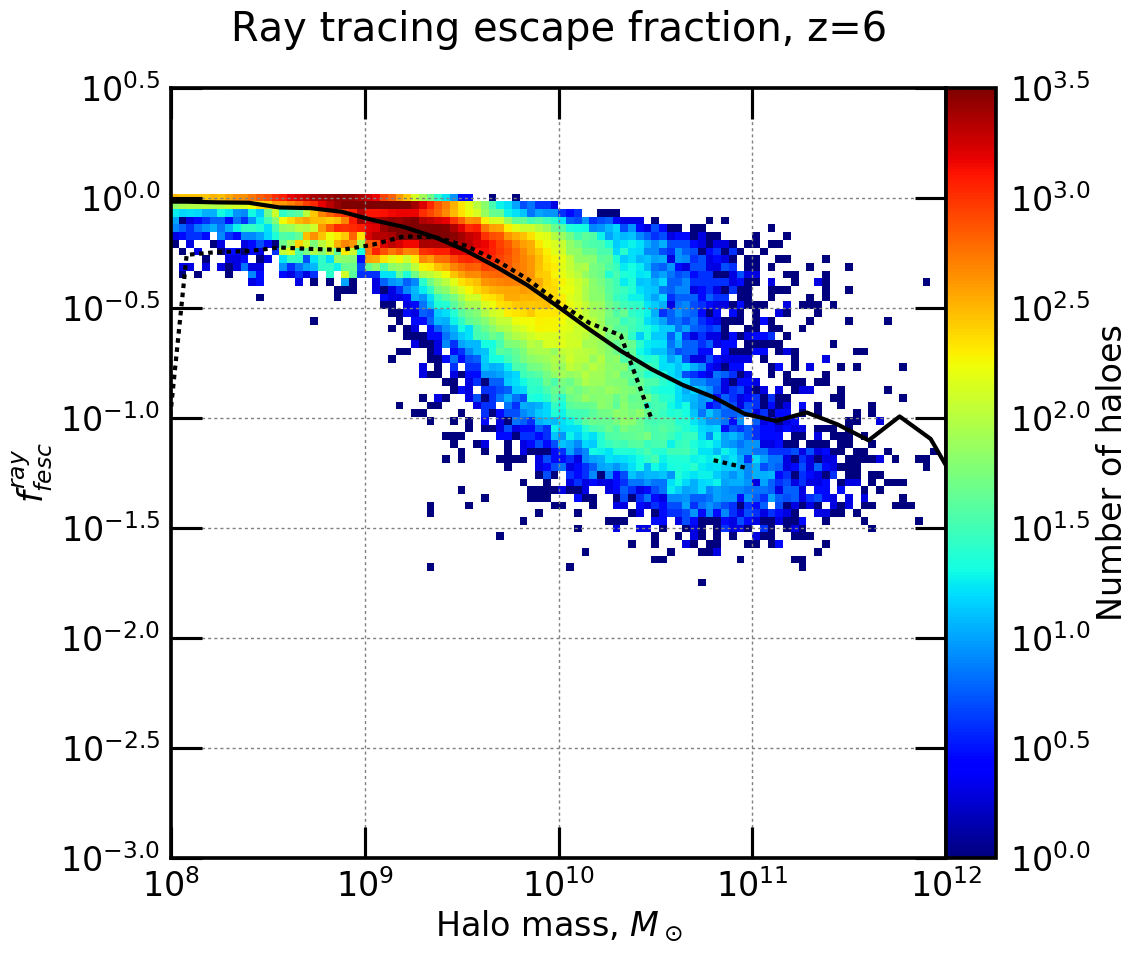}

\caption{Distributions of halo escape fraction $\fescr$ of star-forming haloes as a function of $\mh$ at epochs z=10, 8, 7, 6. The full lines indicate the averages of the distributions, i.e. for the star-forming haloes, whereas the the dotted line indicates averages computed for the whole population, for the z=6 epoch only. The net halo escape fraction can be obtained by multiplying by the sub-grid escape fraction $\fescnet$ =  0.42 $\fescr$. }
\label{fig:escapes}
\end{figure*}

Fig. \ref{fig:escapes} shows the distribution of halo escape fraction $\fescr$ as a function of halo mass for four redshifts : z=6, 7, 8, 10.  

\begin{figure}
    \includegraphics[trim={0cm 0cm 0.cm 0cm},clip,width=\columnwidth]{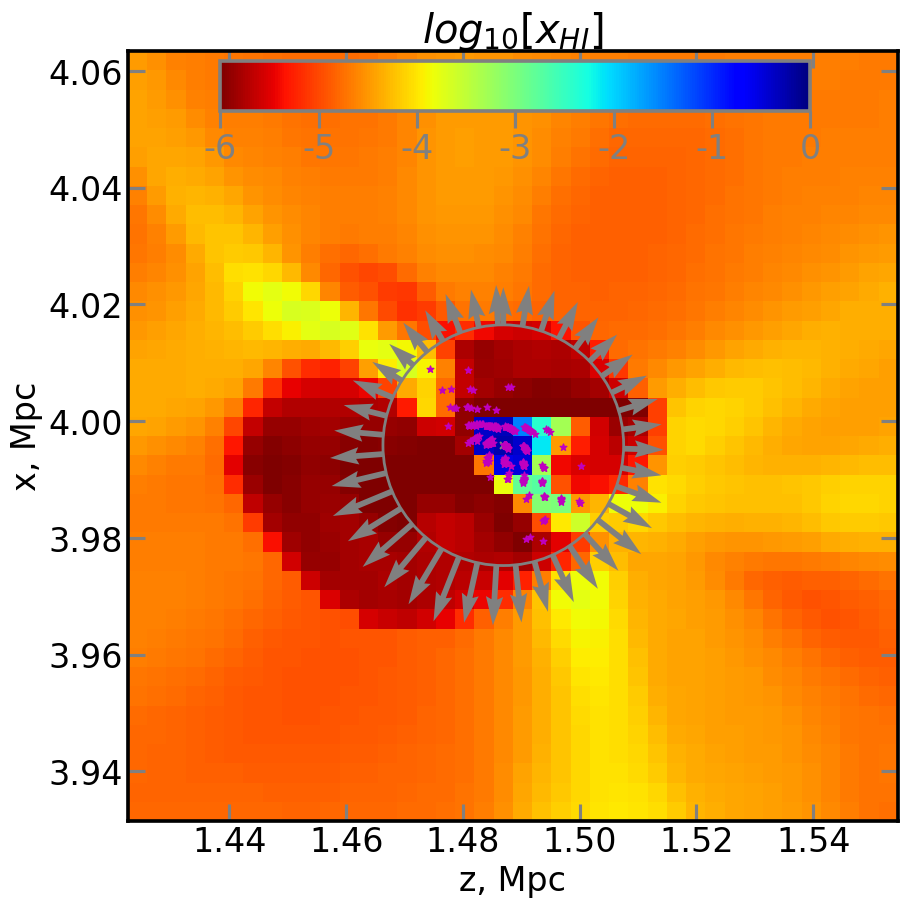}
    \caption{$\xhi$ (colour) centred in a plane containing an example high mass halo ($\approx 5.10^{10} \msol$) at $z=6$. Pink symbols denote active (younger than 10 Myr) stellar particles. Notice the central, neutral region (Several cells are close to fully neutral) that exhibits high star formation. The surrounding ionised region is super heated ($T >> 10^5 K$) and is generated by supernova explosions and accretion shocks. The circle represents $\rtwo$. The gas in the halo centre appears to be connected to several filament-like structures (with $10^{-4.5} \gtrsim \xhi \gtrsim 10^{-3}$).}
    \label{fig:BFH_halo}
  \end{figure}

For masses $\gtrsim 10^{9.5} \msol$, the average $\fescr$ decreases, reaching values of $<10$\% for $\mh > 10^{11} \msol$. These smaller values are observed despite massive haloes appearing to heat their $\rtwo$ sphere surroundings with SN activity, bringing their neutral fraction down to $10^{-6}$ and below. Indeed, massive haloes tend to feature dense, neutral, and opaque cores which trap a large fraction of their ionising photon production as seen in Fig. \ref{fig:BFH_halo}. A more detailed investigation of the internal properties of these haloes is performed in Sec. \ref{sec:drive}.

\begin{figure}
    \includegraphics[trim={0cm 0cm 0.cm 0cm},clip,width=\columnwidth]{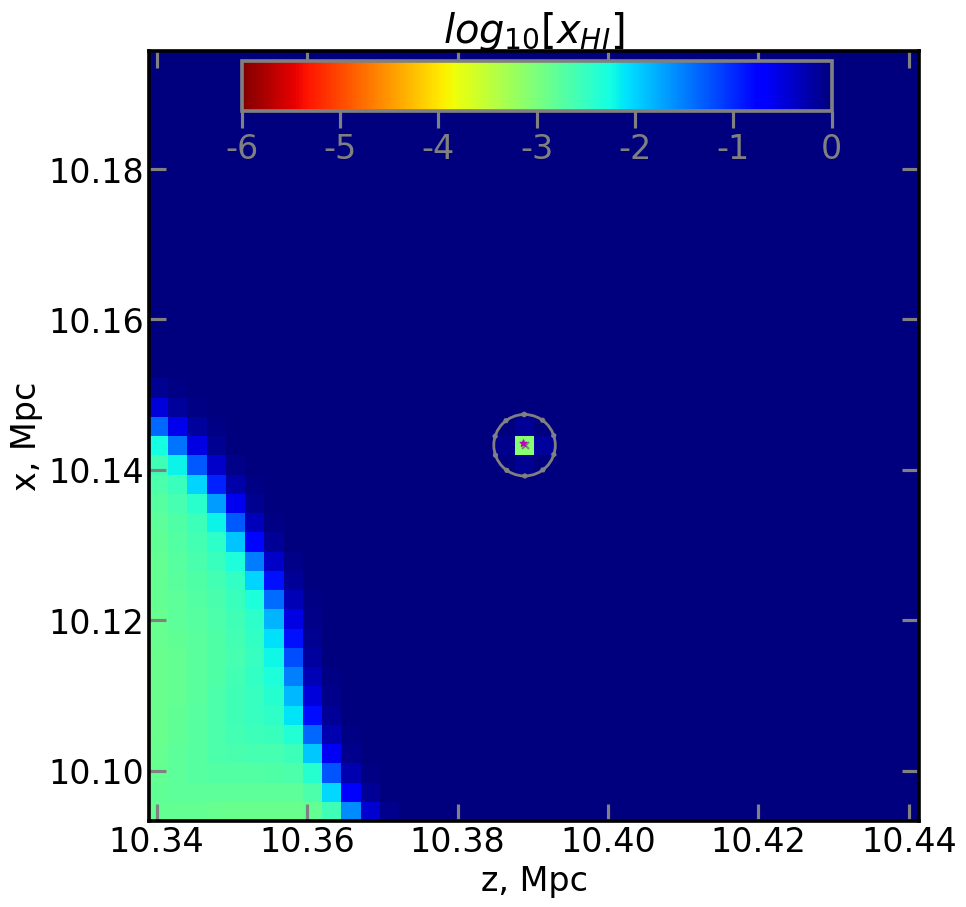}
    \caption{$\xhi$ map (coded by colour) centred on a plane containing an example halo that has just formed a stellar particle at $z=8$. Ionising radiation hasn't yet reached the surfaces where its flux is sampled. Notice the $\xhi$ gradient generated by an incoming UV front in the lower left. The symbols are as in Fig .\ref{fig:BFH_halo}. 
    }
    \label{fig:quies_halo}
  \end{figure}

Halo escape fractions are generally higher at low masses, saturating at 1 for all redshifts. The elongated vertical feature of the distribution of haloes at $\mh \approx 9\times 10^8 \msol$, with $\fesc$ values lower than $10^{-1}$, and at z=10, 8, 7 (top left, top right, bottom left panels of Fig. \ref{fig:escapes}) is populated by haloes in which a star has formed recently, and in which the haloes' expanding HII regions have not yet fully reached the halo boundary at distance $\rtwo$. Fig. \ref{fig:quies_halo} shows the neutral fraction in a plane containing such a halo, illustrating this case. 

Comparing the panels of Fig. \ref{fig:escapes} shows that the extended vertical distribution attributed to haloes in which stars have recently formed progressively disappears between z=10 and z=6, owing to the ionisation of the material of the lower mass haloes by the combination of local UV production, supernovae energy injection, and outside radiation affecting these haloes, thereby rendering them more and more UV transparent within $\rtwo$.

There is an intrinsically high scatter in the escape fractions. It is due to the wide range spanned by the properties of the halo population such as the maximum central density, the gas density profile, and the individual accretion history of the halo. These different properties are well sampled thanks to CoDa II very large size and therefore abundant halo population.

Close examination of the density maps reveals discontinuities around $5\times 10^8 \msol$ and $10^9 \msol$ in the highly populated red/orange areas. These are due to resolution effects. In the case of our less massive haloes, the number of cells that represent them can be small, therefore increases in $\rtwo$ can affect the resulting $\fescr$.

\begin{figure}

  \includegraphics[width=0.9\columnwidth]{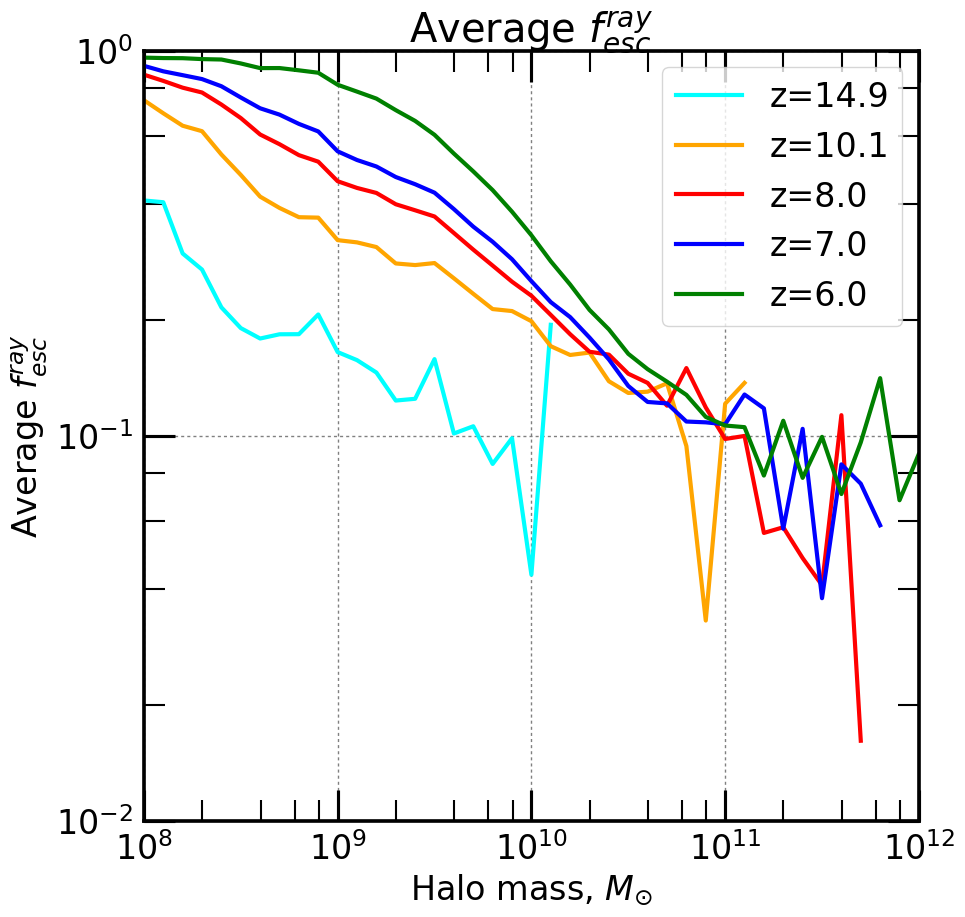}

\caption{Evolution of the average halo escape fraction $\fescr$ of star-forming haloes as a function of $\mh$ and redshift. The net halo escape fraction is obtained by multiplying by the sub-grid escape fraction $\fescnet$ =  0.42 $\fescr$.}
\label{fig:escapes_SFR_z}
\end{figure}

Fig. \ref{fig:escapes_SFR_z}, shows the average escape fractions as a function of mass for five epochs, z=6, 7, 8, 10, 14.9. 
Again, we see high $\fescr$ values for low mass haloes, and a negative mass trend from $\approx 10^9 \msol$ onwards.
There is a clear evolution with redshift for haloes $\mh \lesssim 10^{10} \msol$ : the $\fescr$ average increases with decreasing redshift, reaching $\fescr \approx 1$ at z=6. Indeed, for a fixed halo mass, higher redshift haloes are denser than their lower redshift counterparts, leading to lower escape fractions. The average behaviour of $\fescr$ for haloes $\gtrsim 10^{11} \msol$ is unclear as the number of such objects is small.

\subsection{Comparison with the literature}

The most prominent feature of Fig. \ref{fig:escapes} is a decrease of escape fraction with mass, in agreement with the literature investigating the escape fraction of ionising radiation in high redshift galaxies

from numerical simulations \citep{razoumov_ionizing_2010,yajima_escape_2011,wise_birth_2014,paardekooper_first_2015,katz_census_2018,kimm_escape_2014}, although the slope and extent of the decrease may vary, as can be expected given the range in resolution and modelling of these studies. For dark matter haloes of $10^{11} \msol$, the simulations of \cite{yoo_origin_2020} yield an escape fraction of 7\%, also in rather good agreement with our results.

The other striking feature of our results is the evolution of halo escape fraction with time, in particular at masses below $10^{10}$ $\msol$, showing that in CoDa II, such star forming haloes tend to be more opaque at higher redshifts. For a given halo mass, haloes tend to be denser at higher redshift, and are therefore more likely to yield higher optical depths. { Similar evolution, with escape fractions decreasing at higher redshifts, is seen in \cite{kimm_escape_2014}. \citet{razoumov_ionizing_2010} report an opposite evolution, perhaps due to their modelling of dust (in their study, dust UV optical depth scales linearly with cell density and metallicity, the latter can be expected to increase on average over time, increasing CGM absorption). This may explain their different result, since dust  is not accounted for in our work or in \citet{kimm_escape_2014}.}

\subsection{What drives the decrease of escape fractions with mass?}

\label{sec:drive}

\begin{figure*}

  \includegraphics[width=0.82\textwidth]{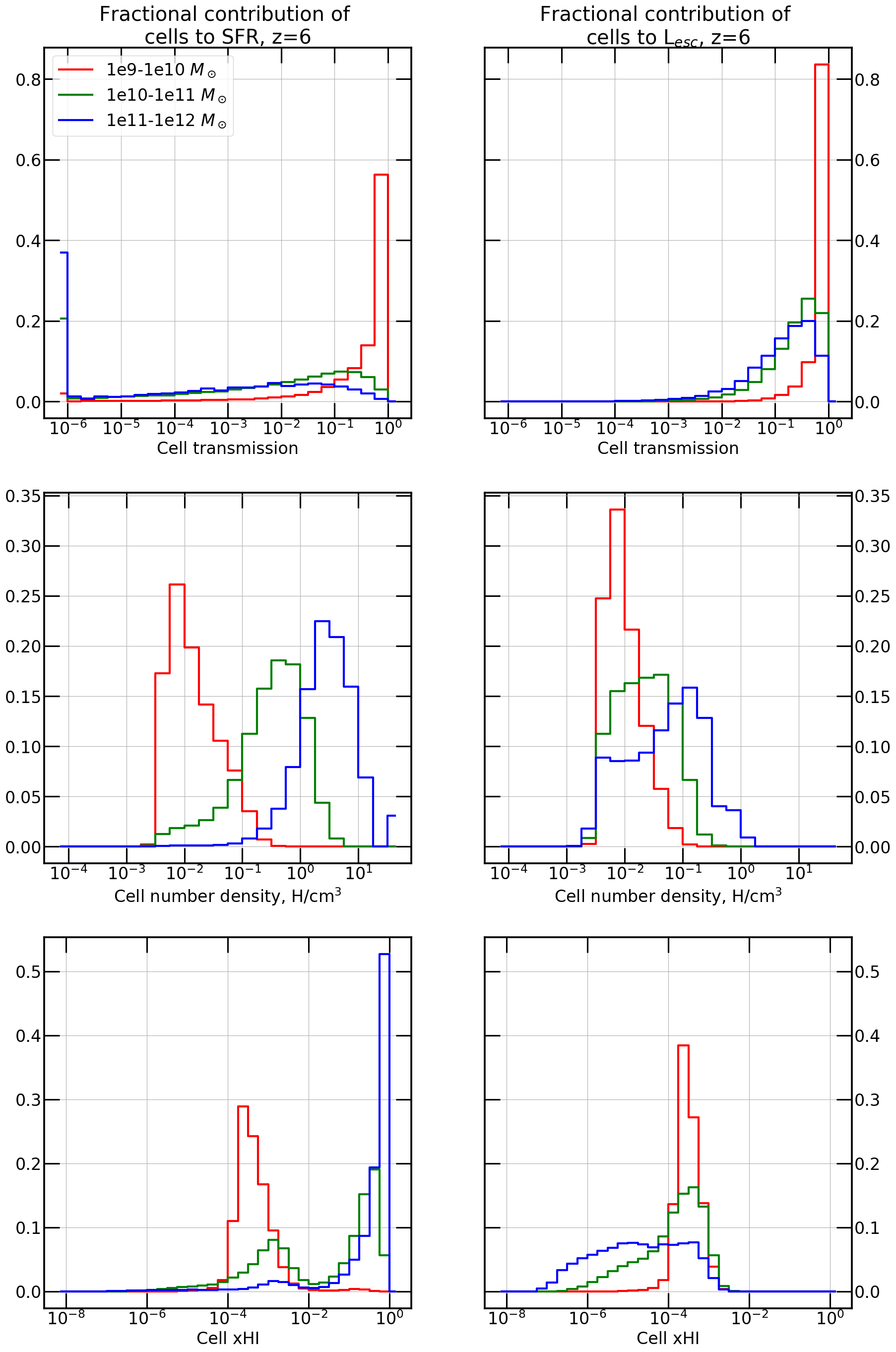}

  \caption{Distribution of star forming halo cell properties: transmissions (top row), physical gas densities in H/cm$^3$(middle row), neutral fractions (bottom row), for three different mass bins of haloes. The histograms show the contribution of halo cells to ({\em left:}) total SFR of the mass bin considered, and to ({\em right:}) the total escape luminosity of the mass bin considered.}
  \label{fig:hist_SFR_pannel}

\end{figure*}

The main trend in all escape fraction plots is a decrease with increasing mass. 
Since halo escape fraction is determined by density and neutral hydrogen fraction, we want to investigate these particular properties.
We show the distributions of the gas properties of star forming cells in Fig. \ref{fig:hist_SFR_pannel}, and their contribution to the SFR and escape luminosity of their halo mass bins.

The top left panel of Fig. \ref{fig:hist_SFR_pannel} shows the distribution of cell transmissions weighted by their SFR for three representative mass bins, normalised by the total SFR in each mass bin (so that the integral of each histogram is 1). 
The star forming cells of the lowest mass bin are concentrated around high transmission values, as observed in individual haloes and yielding the low mass plateaus of $\fescr$ values close to 1. 

There is a stark difference with the intermediate mass bin, where star-forming  cells present a much wider spread of transmission values. The distribution is dominated by cells with intermediate opacity, and peaks at a transmission of $\sim 0.18$.
For the most massive mass bin, however, the dispersion in transmissions is even higher. Between a few times 0.1 and 10$^{-2}$ the distribution is almost flat, and it extends to extremely opaque cells with transmissions as low as $10^{-6}$ and below. The high value of the distribution at $10^{-6}$ indicates that there are close to 40\% of star-forming cells with a transmission lower or equal to $10^{-6}$. There are very few star forming cells with a transmission of 1. This implies that most of the absorption of UV photons within our most massive haloes occurs within the cells containing the sources.
We remind the reader that this is the distribution of the transmission of halo {\em cells}, i.e. 1 value per cell, and not the distribution of the halo-averaged (which would yield 1 value per halo). This is why the distributions extend much lower than in Fig. \ref{fig:escapes}, which show only halo-averaged values. Please note that the peak at the low transmission end is caused by binning all values $<10^{-6}$ together.

We recompute this histogram, weighing this time by the cells' escape luminosity \Lesc. The result is shown in the top right panel of Fig. \ref{fig:hist_SFR_pannel}. This allows us to quantify the contribution of cells to the total escape luminosity of haloes. Unsurprisingly, low mass haloes' escape luminosities originate from high transmission cells. For intermediate and high mass haloes, the situation is slightly more contrasted: while most of the escaping photons originate from high transmission cells, a small fraction (10-15\%) are actually produced by moderately opaque regions with transmission below 0.1.

In order to gain deeper insight into the properties of UV-bright and UV-dark cells, we further examine their physical properties.

The middle left panel of Fig. \ref{fig:hist_SFR_pannel} shows the distribution of cells' gas density for our three representative mass bins, weighted by their SFR, and normalised by the total SFR in each mass bin. The mode of the distribution shifts to higher densities with increasing halo mass, and the density of star-forming cells in the high mass bin extends up to a few 10 H/cm$^3$. This is in stark contrast with the right panel of Fig. \ref{fig:hist_SFR_pannel}, which shows the same distribution, but this time weighted by the cells' escape luminosity \Lesc. The mode of the distributions for the highest and intermediate mass bins are located at much lower densities, and extend no further than about 1 H/cm$^3$, therefore showing that in CoDa II, all stars forming at a density larger than this do not contribute to the escape luminosity of their host halo because their cell transmission is too low.

The bottom left panel of Fig. \ref{fig:hist_SFR_pannel} shows the distribution of cells' neutral hydrogen fractions for our three representative mass bins, weighted by their SFR, and normalised by the total SFR in each mass bin.
For the lowest mass bin, the neutral fraction distribution has one strong peak at just over 10$^{-4}$, which explains their high transmission. The value of $\approx 10^{-4}$ reflects the ionisation equilibrium for a cell of CoDa II resolution at z=6, with an over-density of $\delta \approx 50$ that contains a single emitting stellar particle.
In the case of the $10^{10-11}\msol$ mass bin, haloes' cells are split more or less evenly between ionised and neutral, with two peaks, one centred around 10$^{-3}$ and the other one around a few times 10$^{-1}$ in neutral fractions. 
A possible origin for the binary nature of the distribution may be the rapidity with which dense star forming cells ionise and recombine. In this context, cells would jump very quickly from one peak to the other, depending on their star formation activity or lack thereof, and spend very little time between the peaks.
The star forming cells of the highest mass bin are neutral in majority. Though the distribution presents a small ionised peak as well, the high neutral fraction peak is hugely dominant (about 90$\%$ of the SFR takes place there) when compared to the ionised peak. 

The bottom right panel of Fig. \ref{fig:hist_SFR_pannel} shows the same distribution, but this time weighted by the cells' escape luminosity \Lesc. It shows the contribution of the cells in terms of escaping photons. For low mass haloes, the uni-modal distribution is almost unchanged with respect to the left panel. However, for the two higher mass bins, this different weighing has a strong impact on the relative modes of the distribution: indeed, the high neutral fraction mode of the distribution has completely vanished, so that only the high ionisation mode remains ($\xhi=10^{-3}$ and below), showing that for these high halo masses, escaping ionising photons originate predominantly from strongly ionised regions.
Another striking aspect of the bottom right panel Fig. \ref{fig:hist_SFR_pannel} is the existence of a tail of the distribution, extending to cells with very high ionisation (neutral fractions as as low as $10^{-4}-10^{-7}$) and therefore completely transparent. This tail is absent for the low mass bin.
By examining maps of the physical properties of haloes, such as Fig. \ref{fig:BFH_halo}, we found that such high ionisation within high mass haloes is typical of shocked regions. These shocks can either be accretion shocks as seen in \cite{ocvirk_bimodal_2008,ocvirk_cosmic_2016}, or shocks due to supernovae explosions, or a combination of both. Indeed, several studies have shown supernova feedback to play an important role in the escape of ionising photons \citep{trebitsch_fluctuating_2017,kimm_escape_2014}, and we interpret this very high ionisation tail in our distribution as another manifestation of this effect.
Examining Fig. \ref{fig:BFH_halo} in the light of these results also allows us to locate the regions contributing to the escape luminosity of the halo. The cells of the low ionisation peak, at $\xhi \geq 0.1$, belong to the opaque neutral central core of the halo, from which no ionising photons escape. The high ionisation mode at  $\xhi \sim 10^{-3}$ and below, consists of cells located along the accreting gas filaments, up to the virial radius. Finally, the star forming cells in the very strong ionisation tail of the distribution, at $\xhi<10^{-5}-10^{-6}$ are located in regions heated by supernova feedback.

{
Although the shock-heated tail is not seen in the low mass bins, it shows up when increasing spatial resolution, as demonstrated in Appendix \ref{sec:res_study}. However, it does not result in an increase in $\fesc$, because neutral cells also become more frequent within these haloes, and the net result of increasing resolution is a decrease of $\fesc$.
}

\subsection{Escape luminosities}
\begin{figure}

\includegraphics[width=0.95\columnwidth]{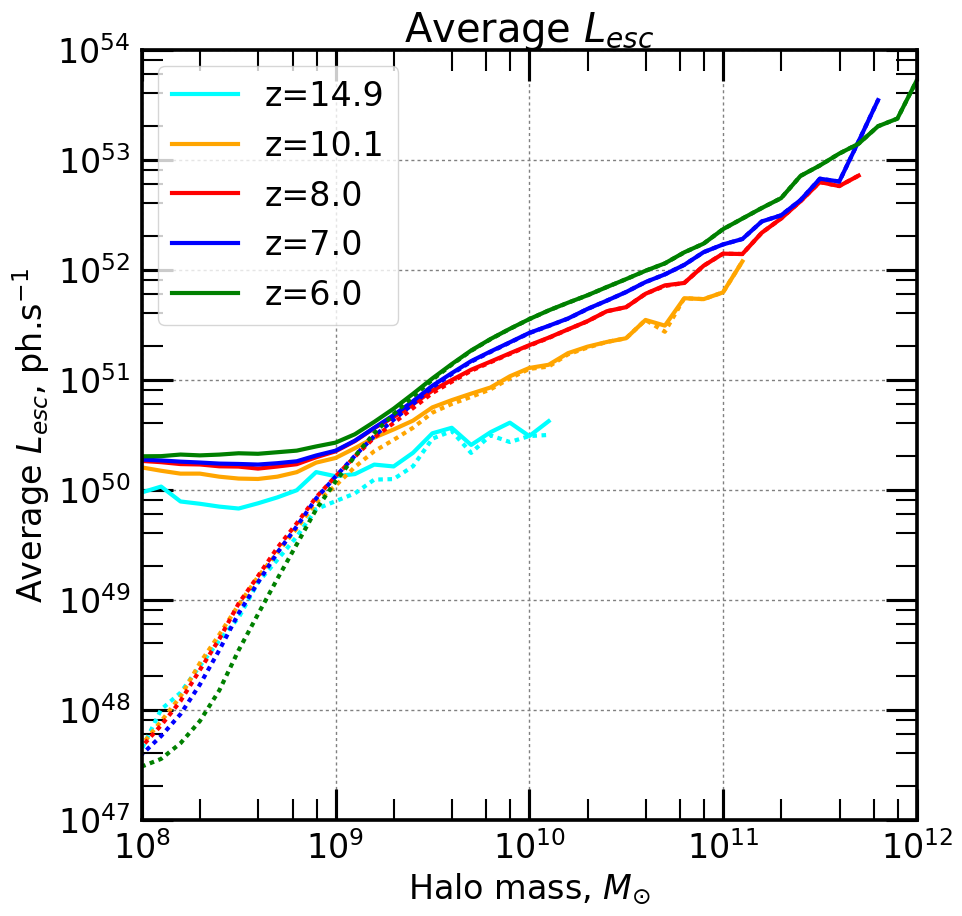}
\caption{Evolution of the average halo escape luminosity \Lesc of star-forming haloes (solid line) and all haloes (dotted line) as a function of $\mh$ and redshift.}
\label{fig:escapes_L_z}
\end{figure}

Using our determinations of halo escape fractions we can now compute the halo escape luminosities as in Eq. \ref{eq:Lesc}. We show the resulting \Lesc as a function of halo mass and their evolution with redshift in Fig. \ref{fig:escapes_L_z}. The escape luminosity increases with halo mass, despite the decrease of $\fesc$. Indeed, at z=6 for instance, $\fesc$ decreases roughly as $\mh^{-0.34} \sim \mh^{-1/3}$, as shown in Appendix \ref{sec:fescfit}, whereas halo SFR increases as $\mh^{5/3}$, and therefore the product \Lesc $\propto$ $\mh^{4/3}$ increases with halo mass. At the low mass end, \Lesc flattens as we reach the quantization limit of star-formation in the simulation: between $10^8-10^9 \msol$, star-forming haloes host only one emitting star particle of the same elementary mass $M_\star = 11732 \msol$. 
The evolution of \Lesc with redshift reflects that of the escape fraction $\fesc$. The impact of star formation suppression by radiation feedback is not readily seen in the solid lines of the figure because they represent the average \Lesc of star-forming haloes only. However, the dotted lines shows the average \Lesc for all haloes, where the suppression-driven decrease of SFR with redshift leads to lower \Lesc at low redshifts below $10^9\msol$.

\section{Results: photon budget}
\label{sec:budget}
We now turn to investigating the contribution of haloes of different masses and different \magn to cosmic reionization in CoDa II. 

\subsection{Photon budget versus mass}

We sort the CoDa II dark matter halo population into 40 logarithmic mass bins between $10^8$ $\msol$ and $10^{12}$ $\msol$. 
We define the total escape luminosity of a given mass bin as the sum of the escape luminosity of all haloes within that mass bin. This quantity depends on the total number of haloes in the CoDa II volume, and we wish to conduct our study using a quantity independent of simulation box size, to ease comparison with future semi-analytical models and simulations. Hence, we further define the escape emissivity, as the total escape luminosity of a given mass bin divided by the simulation volume.

\begin{figure*}
  
\includegraphics[width=\columnwidth]{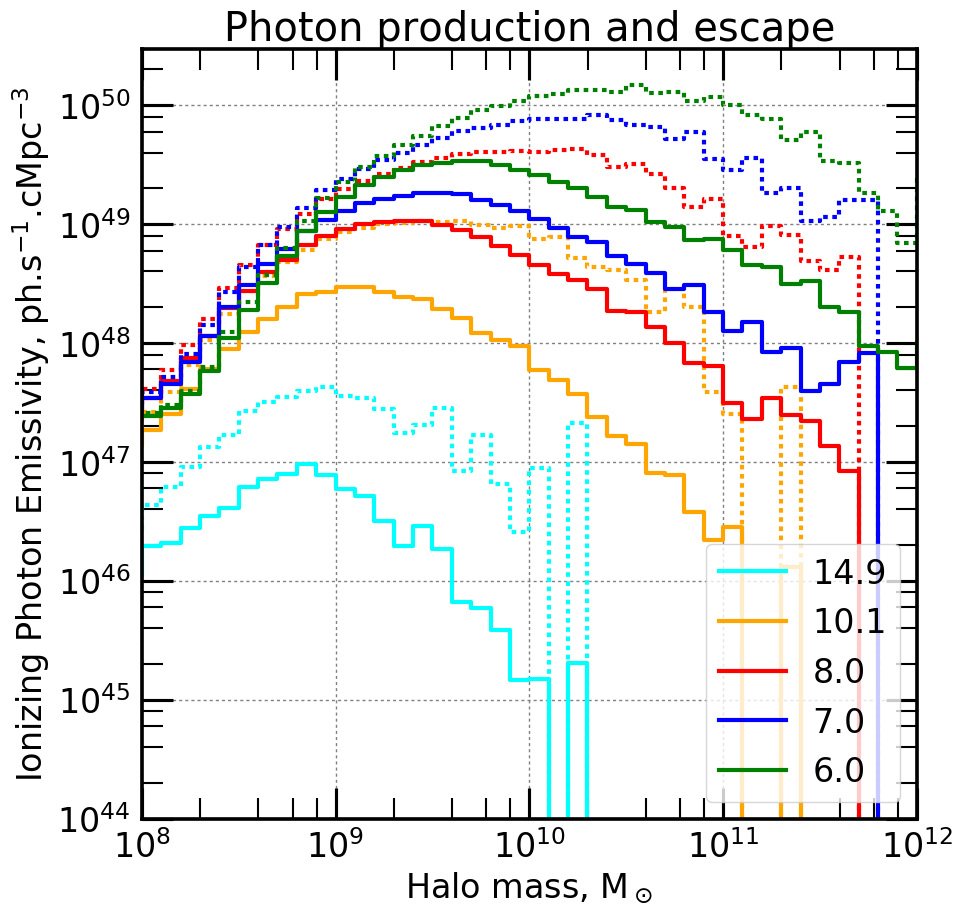}
\includegraphics[width=\columnwidth]{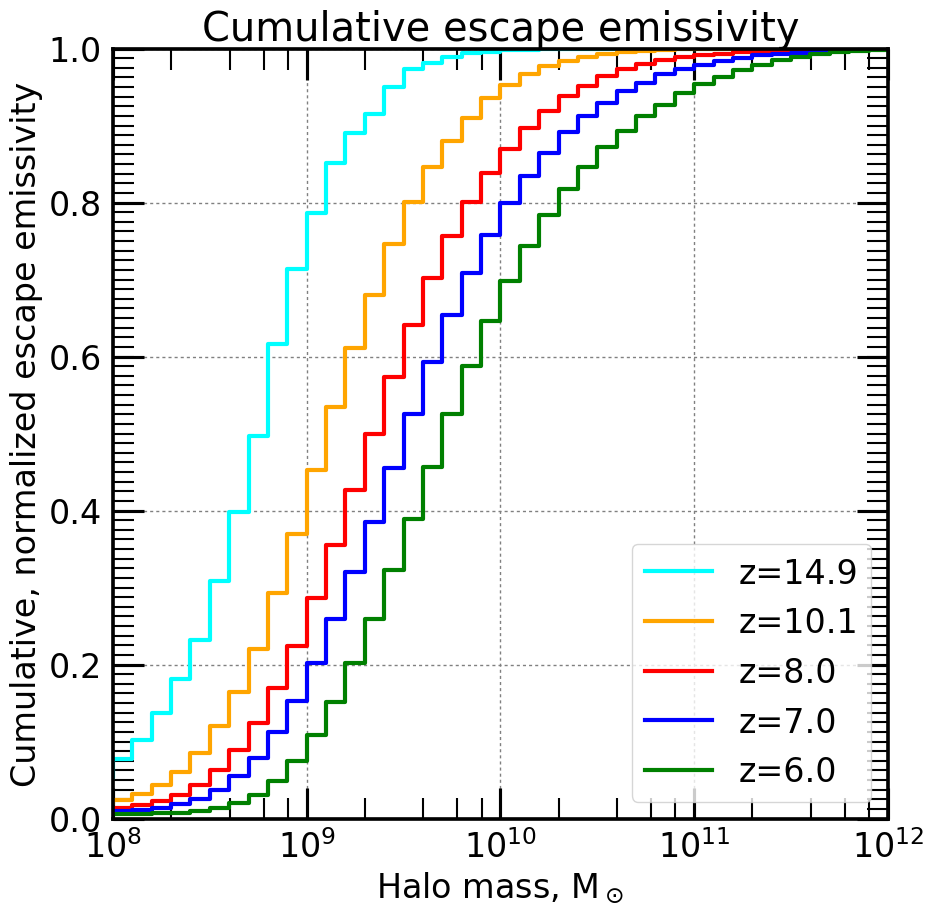}
   \caption{Left : escape emissivity as a function of halo mass for several CoDa II \ epochs. The displayed emissivity is the sum of all haloes' escape luminosities in the mass bin considered at that redshift, divided by the simulation volume. There are 10 mass bins per mass decade. Full lines correspond to the escape emissivity whereas dashed lines correspond to the intrinsic (unabsorbed) emissivities (i.e. derived directly from the star formation rate). Right : Cumulative version of left panel.}
    \label{fig:cii_tot_em}
  \end{figure*}

The left panel of Fig. \ref{fig:cii_tot_em} shows this escape emissivity as a function of halo mass, at 5 epochs (full lines). This represents the cosmic ionising photon budget for the CoDa II simulation, i.e. the distribution of the contributions of each mass bin to the total rate of photons reaching the IGM and driving reionization, for a 1 cMpc$^3$ volume.
 
The contribution to the cosmic escape emissivity culminates around $\sim 10^{9.5} \msol$ between z=6 and z=8 and decreases at lower and higher masses. This leads to a photon budget that is dominated by haloes of a few times $10^8$ up to a few times $10^{10} \msol$ between these redshifts.

In order to quantify the impact of the smaller escape fraction of massive haloes on the photon budget, we also show in the left panel of Fig. \ref{fig:cii_tot_em} the intrinsic photon {\em production} budget, i.e. proportional to the SFR of haloes (dotted line), as compared to the escaping photon budget. At z=6, for instance, intrinsic photon production peaks at $10^{10.5} \msol$. This is the result of two opposite trends: SFR increases with increasing halo mass, whereas the number abundance of haloes decreases, as dictated by the halo mass function. This competition yields $10^{10.5} \msol$ haloes as the foremost contributors to the total cosmic star formation at z=6. However, their contribution to the escaping photon budget is strongly affected by their low escape fractions, to the point that, even though they dominate all other haloes in terms of SFR, they are out-shined in total escape emissivity by the $10^9-10^{10} \msol$ halo population.

To further compare the contribution of various mass bins, the right panel of Fig. \ref{fig:cii_tot_em} shows the cumulative version of the photon budget.
It allows us to directly read from the plot that at z=6, for instance, haloes within $2\times 10^{9} \msol \lesssim \mh \lesssim 10^{10} \msol$ produce around $50 \%$ of all ionising photons. 
The shape of the cumulative photon budget is rather similar at all redshifts.

However, the cumulative distributions shift towards lower masses as redshift increases, by about half a decade between z=6 and z=10. This is also seen as a shift of the peak of the distribution (left panel) between z=6 and z=8. This shift reflects the buildup and evolution of the halo mass function towards more abundant and more massive halo populations.
At all redshifts but the highest, the cumulative distribution has its 10\% and 90\% levels separated by about 1.5 decades in mass, meaning this range is responsible for 80\% of the ionising photon budget. In CoDa II, the ionised volume fraction goes from 20\% to 100\%  between z=6 and 8, and it is $\sim 50\%$ at z=7.
We therefore retain z=7 as the epoch most representative of "ongoing" reionization and detail the photon budget for this redshift: we read from the cumulative distribution that the  photon budget at this epoch is dominated by galaxies with dark matter masses of $6\times 10^8 - 3\times 10^{10} \msol$, which produce 80\% of the ionising photons reaching the IGM. They are therefore the main drivers of cosmic reionization.

\subsection{Photon budget by \magn}

The luminosity function of CoDa II haloes was shown to be in good agreement with observations in \citet{ocvirk_cosmic_2018}. Here we use the halo magnitudes to recast our photon budget analysis into a \magn scale instead of the halo mass scale.

\begin{figure*}
\includegraphics[width=\columnwidth]{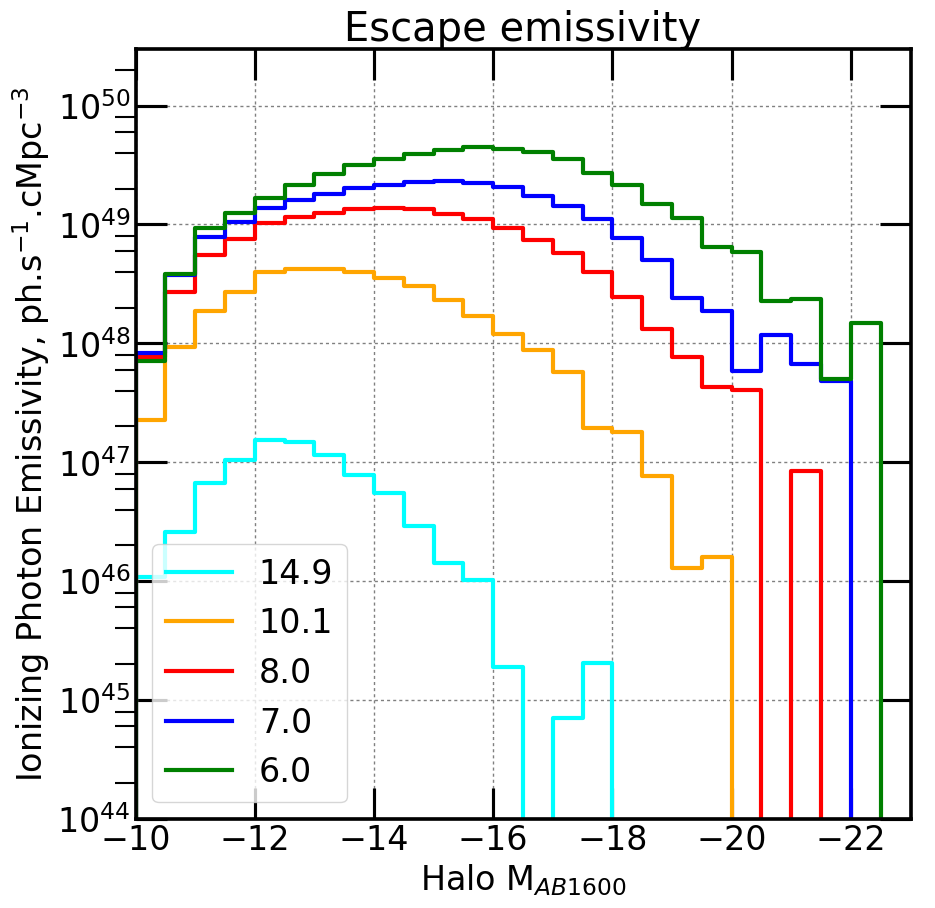}
\includegraphics[width=\columnwidth]{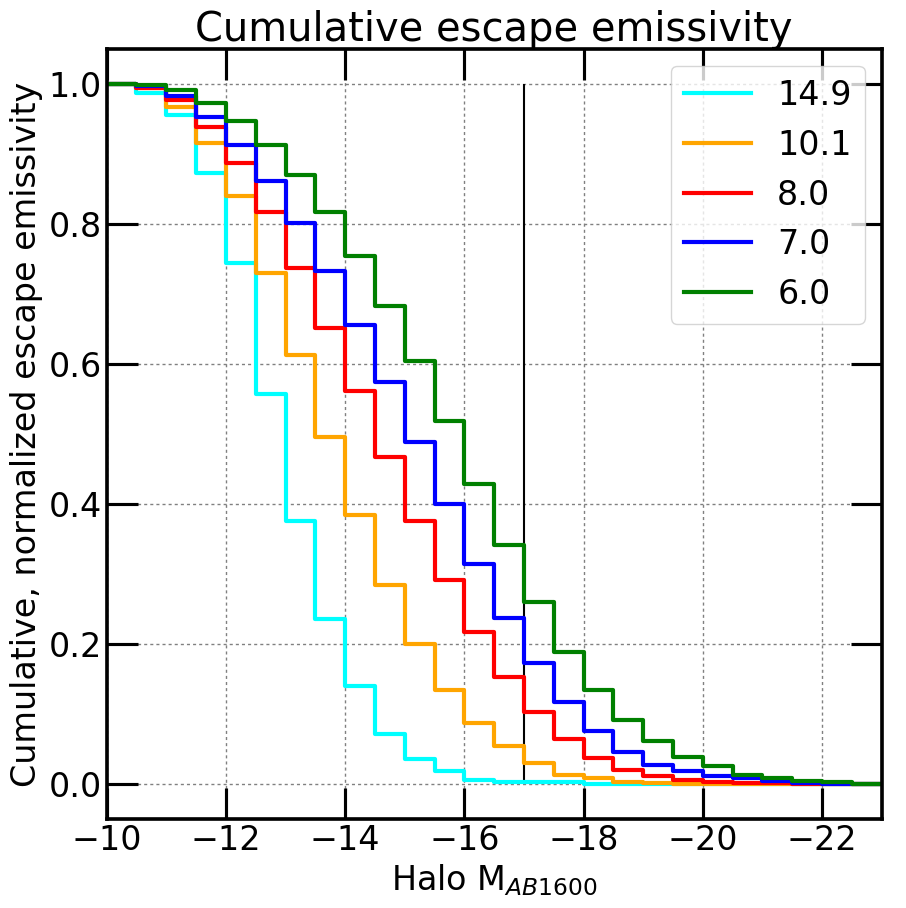}
   \caption{Left : Total escape emissivity as a function of magnitude for several CoDa II \ epochs. The displayed emissivity is the sum of all haloes' escape luminosities in the \magn bin considered, divided by the simulation volume. There are 4 bins per mag. Right : Cumulative version of the total escape emissivity plot on the left. The vertical bar marks the \magn=-17 limit.}
    \label{fig:cii_tot_em_mag}
  \end{figure*}

Fig. \ref{fig:cii_tot_em_mag} shows the total ionising photon contribution as a function of \magn (left), as well as the equivalent cumulative distribution (right). The shape of the distribution and its temporal evolution are similar to that obtained as a function of mass (Fig. \ref{fig:cii_tot_em}). This is a direct consequence of the SFR - halo mass relation shown in \citet{ocvirk_cosmic_2018}, which drives a \magn - halo mass relation.

Combined, these figures illustrate that the main contributors to reionisation lie within a magnitude range of around -12 to about -19. Which is in broad agreement with the semi-analytical results of \citet{liu_dark-ages_2016} for z>7. More specifically, the 80\% escape luminosity range (reading the 10\%-90\% levels of the cumulative escape luminosity distribution function) is \magn = [-13,-19] at z=6, and [-12,-17] at z=8. This shift with redshift is due to the buildup of the galaxy mass function.

This suggests that very deep surveys in cluster fields, such as \cite{bouwens_z_2017} and \cite{atek_extreme_2018}, if reliable down to \magn = -13, are indeed starting to see the bulk (>80\%) of the population driving cosmic reionization at z=6. However, at z=6, cosmic reionization is already finished in CoDa II, and in the middle of reionization, i.e. at z=7, the galaxies seen by  \cite{bouwens_z_2017} are brighter than \magn = -17, and Fig. \ref{fig:cii_tot_em_mag} shows that these can only account for $\sim 20$\% of the ionizing luminosity. In order to see the bulk of galaxies driving reionization when it is in full spin, surveys would need to achieve a similar \magn=-12 depth at z=7. 

Finally, early reionization at z=15 is driven by galaxies of a very narrow range of halo masses, $10^8-10^9 \msol$, corresponding to magnitudes fainter than \magn=-14, out of reach of current and future planned observatories at these redshifts.

\subsection{Comparison with the literature}

\label{ssec:gal_contr}

\citet{katz_census_2018,katz_tracing_2019} use tracers and RAMSES-RT RHD simulations to study the contribution of haloes to the cosmic ionising luminosity. Although their technique differs in many respects from ours (we do not use such tracers, they use variable speed of light, adaptive mesh refinement), their results are in rather good agreement with ours. At z=8, they find that 70\% of the ionising luminosity is produced by haloes of $10^9 \msol \lesssim \mh \lesssim 10^{10} \msol$ \citep[from Fig. 6 of][]{katz_census_2018}. We can read directly from Fig. \ref{fig:cii_tot_em} that at the same redshift, this mass range is responsible for 60\% of the ionising photons in CoDa II. However, at z=6, \cite{katz_census_2018} find that high mass haloes of $\gtrsim 10^{10} \msol$ produce the majority of ionising photons (60\%), while we find that they account for only 40 \% of the ionising luminosity, i.e. the largest contribution originates from haloes less massive than $10^{10} \msol$ in CoDa II.

At higher redshifts, though (z>12), the \cite{katz_census_2018} values fluctuate too much for a meaningful comparison. Also, because of their smaller box size, their sample is devoid of haloes more massive than $10^{11} \msol$ even at z=6, unlike in CoDa II, where such massive haloes are present already at z=10. However, their low number density and escape fractions prevent them from contributing significantly to cosmic reionization: their total escape luminosity is less than 10\% at all redshifts, which is why our results are in fair agreement with \cite{katz_census_2018}, even though they do not include these high mass haloes.

\citet{yajima_escape_2011} find that haloes below $\mh \lesssim 10^{10} \msol$ contribute about $\approx 75$ \% of the ionising luminosity at z=6 (summing the 2 lowest mass bins of the z=6 panel of their Fig. 12). Again, this is in reasonable agreement with our findings, although at this redshift the largest contribution (45\%) comes from haloes in the mass range $10^{9}-10^{9.5} \msol$, while the contribution of this mass range is about 2 times smaller in CoDa II. 
This discrepancy could be due to the lack of radiative feedback on the SFR of low mass haloes in \citep{yajima_escape_2011}: indeed, their study performs RT as post-processing whereas radiation and hydrodynamics are fully coupled in CoDa II, which mitigates star formation in low mass haloes, as shown in \citet{ocvirk_cosmic_2018}, \citet{dawoodbhoy_suppression_2018} and similarly in \citet{wu_simulating_2019,ma_simulating_2018}, and therefore intrinsically reduces their contribution to cosmic reionization.

{ A possible caveat in our work is the assumption we make in Sec. 2.2 : that each halo contains one galaxy. This could potentially result in the blending of galaxies whose real equivalents would be distinguishable in observations as two separate star forming objects. A visual inspection of a sample of our haloes reveals that this preferentially occurs in some of the most massive haloes ($\mh>10^{11}\msol$), at a rate of less than 20\%. Since this mass range only contributes to a few percent of the photon budget, tentatively correcting for this effect  would impact the photon budget only at the percent level, and would not change our conclusions.}

Finally, we address the possible dependence of our results on spatial resolution by performing a follow up higher resolution simulation.

We show in appendix \ref{sec:res_study} that indeed, increasing spatial (mass) resolution in RAMSES-CUDATON by a factor of 4 (64) may yield lower halo escape fractions by a factor of 2, while retaining a similar slope. Such a global re-scaling over all masses, without changing the slope of the $\fesc$ - halo mass relation, should not dramatically affect our results on the photon budget and the predominant halo mass scale driving reionization, because it does not alter the {\em relative} contributions between halo mass bins.

\section{Conclusion}
\label{sec:conclusion}

We use the CoDa II fully coupled RHD simulation of the EoR to study the photon budget of galaxies during the EoR. To do so, we start out by investigating the escape fractions of CoDa II galaxies. We find that the halo escape fraction (i.e. the fraction of ionising photons produced by the halo's stars reaching the virial radius) is a decreasing function of halo mass. 

To gain insight into the evolution of the $\fescr$ with halo mass, we examine the properties of the halo cells as a function of their star formation rates and escape luminosity. We find that for intermediate and high mass haloes, the neutral fractions of star forming cells exhibit a strongly bi-modal distribution, with a neutral mode and an ionised mode at $\xhi \sim 10^{-3}$. The neutral mode is completely opaque, meaning that escaping ionising photons originate from the star forming, ionised regions of the haloes. The halo escape fractions we obtained closely reflect the distribution of the star forming cells between the neutral opaque mode and the ionised, transparent mode. For instance, CoDa II high mass haloes ($10^{11} \msol$) have an average halo escape fraction of  $\sim 10$\% because 90\% of their young stars reside in central, dense, fully opaque regions, while the remaining 10\% of their young stars reside in transparent regions allowing their photons to escape.

Moreover, we find a slow evolution of the halo escape fraction with redshift: haloes of a given mass are more opaque at higher redshift. This is due to the fact that for a fixed mass, haloes at higher redshifts tend to be more concentrated than their low redshift counterparts. 

In Appendix B, we provide a functional form fit to our average halo escape fraction results, so as to allow its use in semi-analytical models of the EoR such as 21cmFAST \citep{mesinger_21cmfast:_2011} or \cite{fialkov_21-cm_2013}.

We then use the halo escape fractions of our haloes to investigate the contributions of galaxies of various masses to the total ionising emissivity during the EoR.
We show that CoDa II  galaxies within $3 \times 10^{10} \msol \gtrsim \mh \gtrsim 6 \times 10^8 \msol$ produce about 80\% of all the ionising photons reaching the IGM at z=7, which is the middle of reionization in CoDa II ($\rm{x_{HI}}=50\%$). 
They can therefore be considered as the main drivers of cosmic reionization, although, 
at z=6 (8), the mass range accounting for 80\% of the photon budget is slightly more (less) massive, by 0.25 dex.

The foremost mass range reionizing the Universe emerges as the result of a competition between the different processes exposed throughout this paper, and can be summarised as follows : the numerous low mass haloes are too inefficient at forming stars to contribute significantly despite their high halo escape fractions, whereas the high mass haloes are too few and have escape fractions that are too low to contribute significantly, despite their high star formation rate.

As a consequence, the low mass end (below $5\times 10^9 \msol$) and the high mass end (above $5 \times 10^{10} \msol$) contribute respectively only less than 10\% each to the total ionising photon budget between z=8 and z=6.

Our results are in reasonable agreement with the (not exhaustive) literature reviewed, despite a number of differences in numerical treatment, and assumptions on stellar populations and their feedback, which explain the deviations from our results.

Ideally, we would like to follow up on our study by pushing future CoDa II-like simulations to higher spatial resolutions, possibly using AMR to provide a better description of the ISM, and its processes and if possible rely less on a sub-grid escape fraction, as well as improved physics such as chemical enrichment, AGN mechanical and radiative feedbacks, and their possible contribution to reionization. The continued growth of supercomputers, thanks to hybrid nodes mixing many-core CPUs and GPUs, may allow us to get there in the near future, provided we overcome a number of technical hurdles related to code architecture and optimisation. 

\section*{Acknowledgements}
The authors would like to acknowledge the judicious comments of the anonymous referee, as well as those of Max Gronke. 
This work made use of Python, and the following packages Python : matplotlib \citep{Hunter:2007}, numpy \citep{van2011numpy}, scipy \citep{Virtanen_2020}, healpix \citep{gorski_HEALPix_2005}.
This study was performed in the context of several French ANR (Agence Nationale de la Recherche) projects. 
PO acknowledges  support  from  the  French  ANR  funded  project ORAGE  (ANR-14-CE33-0016).  
ND  and  DA  acknowledge funding  from  the  French  ANR  for  project  ANR-12-JS05-0001 (EMMA). 
ITI was supported by the Science and Technology Facilities Council (grant numbers ST/F002858/1 and ST/I000976/1) and the Southeast Physics Network (SEPNet).  
JS  acknowledges support from l'Or\'eal-UNESCO "Pour les femmes et la Science” and the "CNES (Centre National d'\'etudes spatiales)" postdoctoral fellowship programs. 
KA was supported by NRF (Grant No. NRF-2016R1D1A1B04935414).
GY acknowledges financial support by the MINECO/FEDER under project grant AYA2015-63810-P and MICIU/FEDER under project grant PGC2018-094975-C21.
PRS was supported in part by U.S. NSF grant AST-1009799, NASA grant NNX11AE09G, NASA/JPL  grant  RSA  Nos.  1492788  and  1515294,  and supercomputer resources from NSF XSEDE grant TG-AST090005  and  the  Texas  Advanced  Computing  Center(TACC)  at  the  University  of  Texas  at  Austin. 
The CoDa II simulation was performed at Oak  Ridge  National  Laboratory/Oak  Ridge  Leadership Computing Facility on the Titan supercomputer (INCITE2016 award AST031). Processing was performed on the Eos and  Rhea  clusters.  Auxiliary  simulations  were  performed at  p\^ole  HPC  de  l’Universit\'e  de  Strasbourg  (m\'eso-centre). The simulations used for the resolution study were performed on
CSCS/Piz Daint (Swiss National Supercomputing Centre), as part of the “SALT: Shining a light through the dark ages” PRACE allocation obtained via the 16th call for PRACE
Project Access (project id pr37).
A  series of test simulations  for the initial conditions of CoDa II were performed at LRZ Munich within the project pr74no.
This work made use of v2.1 of the Binary Population and Spectral Synthesis (BPASS) models as last described in \citet{eldridge_binary_2017}.

\section*{Data Availability}
The data underlying this article will be shared on reasonable request to the corresponding author. See \citet{ocvirk_cosmic_2018} for specifics about the availability of CoDa II.

\appendix

\section{Escape fraction resolution study }
\label{sec:res_study}
We have shown that the main features of our $\fescr$ values are that they decrease with increasing halo mass above a certain mass scale, with haloes of $\lesssim 10^9 \msol$ having high $\fescr$ values, and increase slightly with the progress of the EoR. Since this first trend is driven by the presence of dense, neutral and UV opaque cells in haloes, one may expect that increasing the number of resolution elements (and thus within the same volume and same total mass, increasing the maximum possible density) could affect the exact relation between $\fescr$ and halo mass, as well as the mass scale at which the break between trends occurs.

In order to investigate this, and to test the robustness of our previously presented results, we proceed to study the resolution convergence of $\fescr$.
{ We ran a series of high resolution simulations using RAMSES-CUDATON : we used a 4 cMpc.h$^{-1}$ sided box, with 1024$^3$ resolution elements (ie : 4 times the spatial resolution of CoDa II, i.e. 64 times better mass resolution). We will call it high-res from now on. This setup yields a comoving cell size of 5.76 ckpc (0.82 kpc physical at z=6), and a dark matter particle mass of $6.4 \times 10^{3} \msol$. All other parameters are kept exactly as in CoDa II, including the minimum stellar mass particle of 11732 $\msol$.}
The initial conditions are necessarily different from CoDa II, because the high-res run uses a smaller box. As a consequence, it contains fewer star forming haloes of a given mass than CoDa II, making $\fescr$ measurement noisier. The latter must therefore be carefully compared with oru former results.

We performed our previous $\fescr$ measurements on the high-res simulation in order to test the impact of numerical resolution on $\fescr$ in CoDa II.

\begin{figure*}
    \includegraphics[width=\columnwidth]{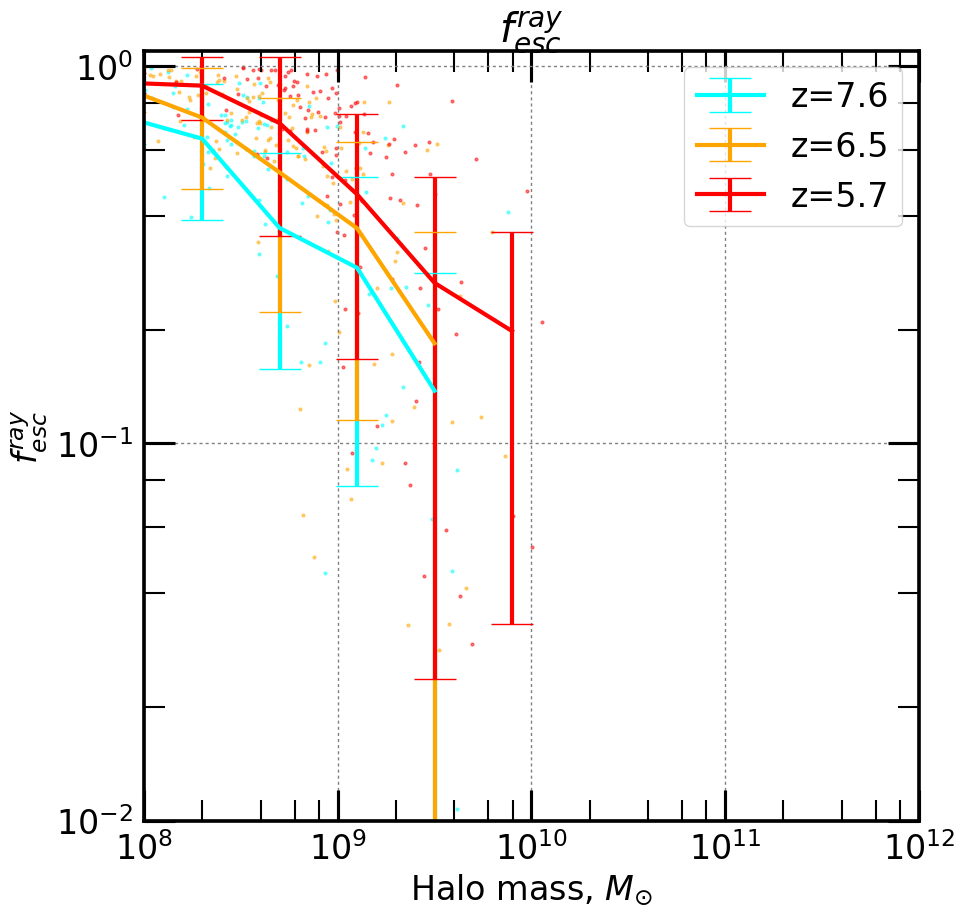}
\includegraphics[width=\columnwidth]{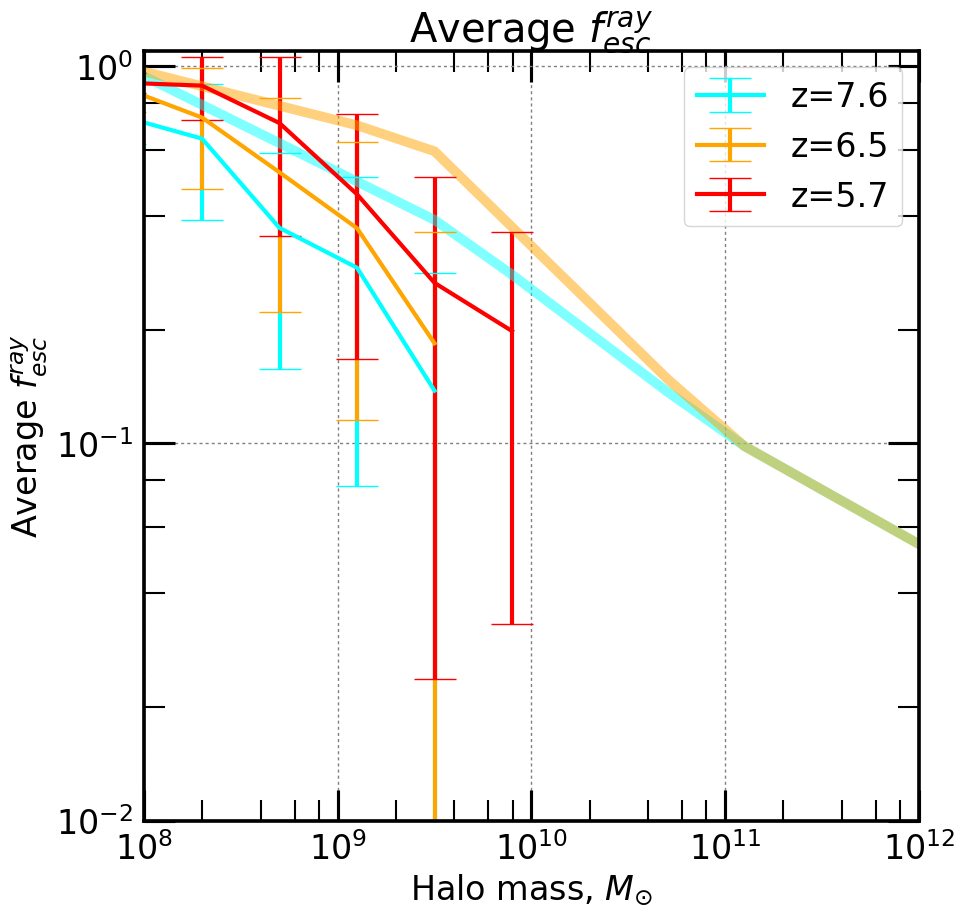}
   \caption{Left : Scatter and mass bin average of $\fescr$ of star-forming haloes over the full range of masses in the high-res simulation box | Right : average $\fescr$ of star-forming haloes for the high-res box, compared with our fitting formula (See Eq. \ref{eq:fesc_fct}) for CoDa II $\fescr$ (at redshifts that are appropriate, 10>z>6). Here we plot within the mass range of CoDa II haloes.}
    \label{fig:fesc_RSL_1024}
  \end{figure*}
  
The left panel of Fig. \ref{fig:fesc_RSL_1024} shows $\fescr$ as a function of halo mass for the high-res simulation. This high resolution case also presents high $\fescr$ values for low mass haloes, as well as decreasing $\fesc$ with halo mass. Moreover, $\fescr$ increases on average with time for low mass haloes, as in CoDa II. The right panel presents a direct comparison of these averages with the fitting formula for CoDa II at the same redshifts where appropriate ($10\gtrsim z \gtrsim 6$, the domain of validity for our fitting formula given in \ref{eq:fesc_fct}).

However, as anticipated, there are differences. In the high-res box, the average $\fescr$ is lower for all masses than in CoDa II,  and the slope with mass is slightly more pronounced. However, there is a large scatter around the average of the high-res data-set. Indeed, there are only a few hundred star forming haloes of all masses in the high-res box at z=5.7.
Reassuringly, the difference between the two boxes is akin to a global re-normalisation of the average $\fescr$. Therefore, while numerical resolution may change the {\em absolute} \Lesc of a given halo mass bin, it is not likely to change the {\em relative} balance between mass bins in the photon budget, which is our main result.

\begin{figure*}

  \includegraphics[width=0.9\textwidth]{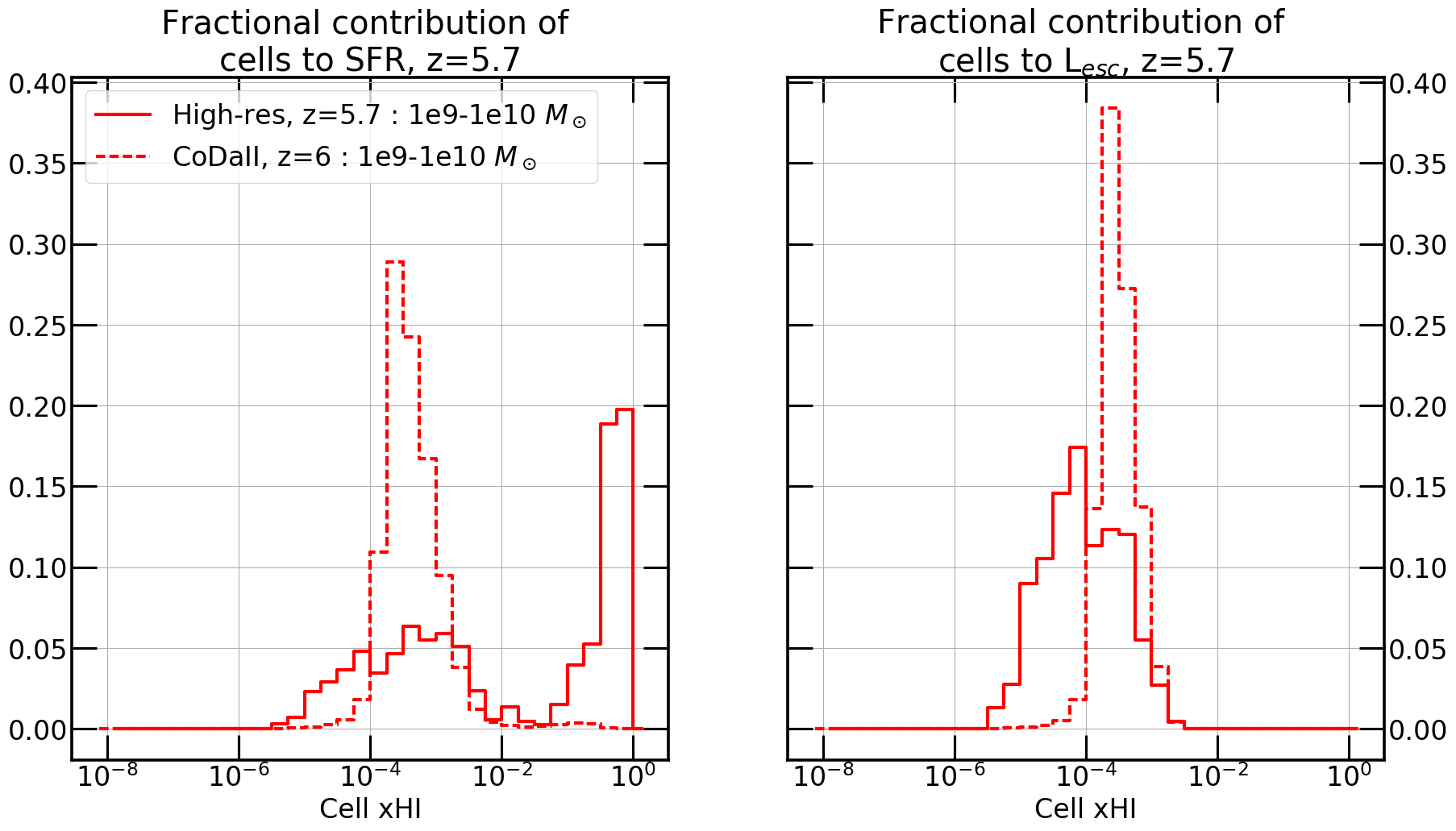}
  \caption{Distribution of star forming halo cell neutral fractions for haloes of masses $10^{10}\msol >  \mh > 10^{9}\msol$. Full lines show the curves for the high-res simulation, whereas dashed lines show the curves for the CoDaII simulation. The histograms show the contribution of halo cells to ({\em left:}) total SFR of the mass bin considered, and to ({\em right:}) the total escape luminosity of the mass bin considered.}
  \label{fig:hist_SFR_pannel_highres}
\end{figure*}

{ In order to explain the smaller $\fescr$ values measured in high-res, we proceed as in \ref{sec:drive}, and examine the properties of the gas of star forming cells.  Fig. \ref{fig:hist_SFR_pannel_highres} shows the distribution of the neutral fractions of star forming cells in high-res, weighted by their SFR (\Lesc) in the left (right) panel, at z=5.7. In high-res, the neutral fractions of star forming cells in haloes of  $10^{10}\msol >  \mh > 10^{9}\msol$ are grouped into two peaks : an ionised mode, centred around $\xhi \sim 10^{-4}$, and a neutral / quasi-neutral mode at $\xhi \sim  0.5$. The \Lesc weighted distribution shows that only the cells with  $ \xhi \leq 10^{-3}$ contribute significantly to the final \Lesc of the haloes in high-res. The cells belonging to the high neutral fraction account for $\approx 50\%$ of star formation, and do not contribute to haloes' \Lesc . 

Fig. \ref{fig:hist_SFR_pannel_highres} also shows the CoDa II distributions, allowing us to gauge directly the impact of increased resolution.
The first striking difference between the two simulations is the shape of the SFR weighted distribution. In CoDa II most of the star formation in the $10^{10}\msol >  \mh > 10^{9}\msol$ haloes happens in cells with neutral fractions <$10^{-3}$, and there is no equivalent to the high neutral fraction peak seen in high-res. However, this feature is present in the distributions of cell xHI in higher mass haloes in CoDa II, that are better resolved. This is a convincing hint that the increase of resolution between CoDa II and high-res allows higher density cells to exist within halos of the same mass, in turn allowing for higher recombination rates, higher neutral fractions, and finally lower LyC transmissions and escape fractions.

There is however, a second aspect worthy of discussion. In CoDa II, the star forming cells of haloes of $10^{10}\msol >  \mh > 10^{9}\msol$ do not exhibit neutral fractions smaller than $10^{-4}$, whereas in high-res the distributions of neutral fraction stretch all the way to $10^{-5}$. Higher mass haloes in CoDa II have similarly low neutral fractions that can only occur when the gas is heated by supernova. This suggests that although the higher resolution of high-res allows for higher gas densities that can decrease the escape fraction, it also permits the existence of lower gas density cells that are more susceptible to the heating from supernova, plausibly increasing the $\fescr$ of haloes by allowing higher transmission cells to exist. The fact that the escape fractions of the high-res box are lower suggests that this transmission boost does not compensate the opacity boost occurring at higher resolution, and the latter dominates. 

}

\section{Escape fraction fits}
\label{sec:fescfit}

In this section, we propose simple functional forms for the both the average of $\fescr$ and the SFR-weighted average of $\fescr$ as a function of halo mass and of redshift. These could be useful as a $\fescr$ model for projects in which the determination of $\fescr$ is impossible or difficult : either when working with simulated data that wasn't produced with fully coupled radiation-hydrodynamics simulation codes, performing radiative transfer in post-processing, or with simulations with lower spatial resolution than CoDa II, and also for semi-analytical models of the EoR \citep{mesinger_21cmfast:_2011,fialkov_21-cm_2013}.

In all cases, we caution the reader that the functions presented here represent the halo escape fraction, i.e. in order to obtain the net halo escape fraction of haloes, $\fescnet$, one needs to multiply $\fescr$ by $\fescsub$ (0.42), as show in Eq. \ref{eq:fesc_def}.

{ We provide the reader with python functions for plotting the fits that follow \href{https://github.com/jlewis47/CoDa-II-escape-fraction-fit-functions/blob/master/CoDaII_avg_fesc_fcts.py}{here (github)}.}

\subsection{Average escape fraction}

Based on the trends presented by the average values of $\fescr$ as a function of mass and of redshift between z=6 and z=10 shown in Fig. \ref{fig:escapes_SFR_z}, we opted for a double power law of halo mass, with a knee separating both laws, that shifts with redshift. 
Due to the noisy appearance of the data for $\rm \mh > 10^{11} \msol$, the time behaviour of $\fescr$ within this mass range is unclear, which is why we chose to simply model the evolution of $\fescr$ with mass as the extension of the  $\rm f^{\rm ray}_{\rm esc}(\mh,z=10)$ fit, in order to produce indicative values despite the high amount of noise in the average curves for these masses.

We propose the following functional form (Eq. \ref{eq:fesc_fct}):

{
\begin{equation}
    \fescr= \left \{
    \begin{array}[]{ll}
    \rm f^{\rm ray, 10.1\leq z\leq 14.9}_{\rm esc}(\mh,z); & \rm if \, \mh>M_{join} \, , \\
    \rm f^{\rm knee}_{\rm esc}(z)f^{\rm knee}(\mh,z); & \rm if \, M_{\rm join} \geq \mh \geq M_{\rm knee}(\rm z) \, ,\\
    \rm f^{\rm max}_{\rm esc}(z)f^{\rm max}(\mh,z); & \rm if \, \mh < M_{\rm knee}(\rm z)\, ,
    \end{array}
    \right.
    \label{eq:fesc_fct}
\end{equation}

\begin{equation}
    \rm with= \left \{
    \begin{array}[]{lcl}
    \rm z_6 & = & \Big(\frac{1+z}{1+6}\Big)\, , \\
    \rm f^{\rm max}_{\rm esc}(z) & = & \rm min\big(\rm f^{\rm 0, max}_{\rm esc} z_6^{\beta},\,1.0 \big) \, , \\
    \rm f^{\rm knee}_{\rm esc}(z) & = & \rm max\Big( min\big(\rm f^{\rm 0, knee}_{\rm esc} z_6^\gamma,\,1.0 \big),f^{\rm knee}_{min}\Big) \, , \\
    \Delta^{\rm max}(z) & = & \frac{\rm log_{10}(\rm f^{\rm knee}_{\rm esc}(z))-log_{10}(\rm f^{\rm max}_{\rm esc}(z))}{\rm log_{10}(M_{\rm knee}(z))-log_{10}(10^8)} \, , \\
    \Delta^{\rm knee}(z) & = & \frac{\rm log_{10}(\rm f^{\rm knee}_{\rm esc}(z))-log_{10}(\rm f^{\rm join}_{\rm esc})}{\rm log_{10}(M_{\rm knee}(z))-log_{10}(M_{\rm join})} \, , \\
    \rm M_{\rm knee}(\rm z) & = &2.5\times 10^9  z_6^\zeta \msol \, , \\
    \rm f^{\rm max}(\mh,z) & = & \Big(\frac{ \mh}{\rm M_{\rm knee}(\rm z)}\Big)^{\Delta^{\rm max}(z)} \, , \\
    \rm f^{\rm knee}(\mh,z) & = & \Big(\frac{ \mh}{\rm M_{\rm knee}(\rm z)}\Big)^{\Delta^{\rm knee}(z)} \, , \\
    \end{array}
    \right.
    \label{eq:fesc_fct_params}
\end{equation}

$\rm f^{\rm max}_{\rm esc}(z)$ is the maximum value below $\rm M_{\rm knee}$, it is $f^{\rm 0, max}_{\rm esc}$ at z=6, and evolves as $z^\beta$.\\
$\rm f^{\rm knee}_{\rm esc}(z)$ is the where both power laws join at $\rm M_{\rm knee}(z)$, it is $f^{\rm 0, knee}_{\rm esc}$ at z=6, and evolves as $z^\gamma$. It cannot be lower than $f^{\rm knee}_{min}$.\\
$\Delta^{\rm max}(z)$ is the slope of the power law above $\rm M_{\rm knee}(z)$, it is defined so as to reach $\rm f^{\rm max}_{\rm esc}(z)$ when M=$10^8\msol$.\\
$\Delta^{\rm knee}(z)$ is the slope of the power law below $\rm M_{\rm knee}(z)$, it is defined so as to reach $\rm f^{\rm knee}_{\rm esc}(z)$ when M=$\rm M_{\rm knee}(z)$, and $\rm f^{\rm join}_{\rm esc}$ when M=$\rm M_{\rm join}$.\\
$\rm M_{\rm knee}(\rm z)$ gives the mass where the power laws join, it is $2.5\times10^9 \msol$ at z=6, and evolves with z as $z^\zeta$.\\
$\rm f^{\rm max}(\mh,z)$ accounts for the mass slope for $\mh<\rm M_{\rm knee}(\rm z)$ given by $\Delta(z)$.\\
$\rm f^{\rm knee}(\mh,z)$ accounts for the mass slope for $\mh \geq \rm M_{\rm knee}(\rm z)$ given by $\delta$.\\}

The adjusted values are summarized in Eq. \ref{eq:fesc_fits_vals} below.
\begin{equation}
    \left\{
    \begin{array}[]{lrr}
    \rm f^{\rm 0, max}_{\rm esc} & = & 0.98 \, , \\
    \rm f^{\rm 0, knee}_{\rm esc} & = & 0.8 \, , \\
    \rm M_{\rm join} & = & 10^{11} \msol \, , \\
    \rm f^{\rm join}_{\rm esc} & = & 0.1 \, , \\
    \rm f^{\rm knee}_{min} & = & 0.23\, , \\
    \beta & = & -0.5 \, , \\
    \gamma & = & -3.5 \, , \\
    \zeta & = & -2.0\, , \\
  \end{array}
  \right.
  \label{eq:fesc_fits_vals}
\end{equation}

The average curves corresponding to 10$\leq$ z $\leq$ 14.9 behave somewhat differently. They appear to be simple power laws, offset in escape fraction by $\sim 0.25$ dex. Hence, instead of using the previously discussed formulas for this redshift range, we tentatively provide the following fit (Eq. \ref{eq:fesc_fct_z15}) for $\fescr$ for 10$\leq$ z $\leq$ 14.9.

\begin{equation}
\rm f^{\rm ray, 10.1\leq z\leq 14.9}_{\rm esc}(\mh,z) = {\rm min} \left( \rm f^{\rm 10\leq z \leq 14.9}_{\rm esc,0}\times\Big(\frac{\mh}{10^8 \msol}\Big)^{\Delta^{10\leq z \leq 14.9}}\times \big(\frac{10.}{z}\big)^{\rm \gamma} \, , 1 \right) \, ,
\label{eq:fesc_fct_z15}
\end{equation}

\begin{equation}
    \rm with= \left \{
    \begin{array}[]{lcl}
    \rm f^{\rm 10.1\leq z\leq14.9}_{\rm esc,0} & = & 0.77\, , \\
    \rm \Delta^{10.1\leq z\leq14.9} & = & \frac{\rm log_{10}(0.77)-log_{10}(\rm f^{\rm join}_{\rm esc})}{\rm log_{10}(10^8 \msol)-log_{10}(M_{\rm join})}\, , \\
    \rm \gamma &=& -1.64 \, , \\
    \end{array}
    \right.
    \label{eq:fesc_fct_z15_params}
\end{equation}

Once again, as we aim to represent the averages and not fit the data, we adjust these values by hand. 

\begin{figure}
\includegraphics[width=\columnwidth]{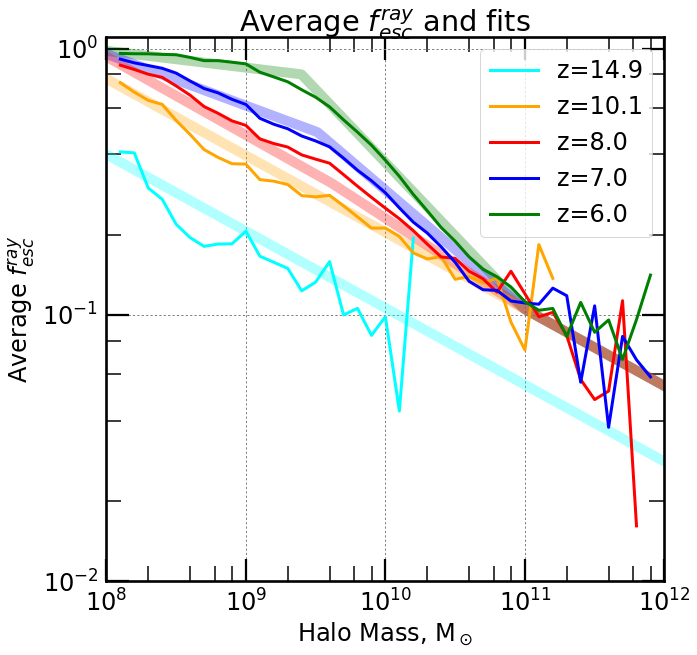}
   \caption{Average escape fractions (thin lines) of star-forming haloes and corresponding fits (thick lines) at z=6, 7, 8, 10.1, 14.9.}
    \label{fig:fesc_fits}
  \end{figure}

The aforementioned fits are presented in Fig. \ref{fig:fesc_fits} (full, thick lines), where they are also compared to the mass bin averages of $\fescr$ (full, thin lines).

\subsection{SFR weighted average escape fraction}

\begin{figure}
\includegraphics[width=\columnwidth]{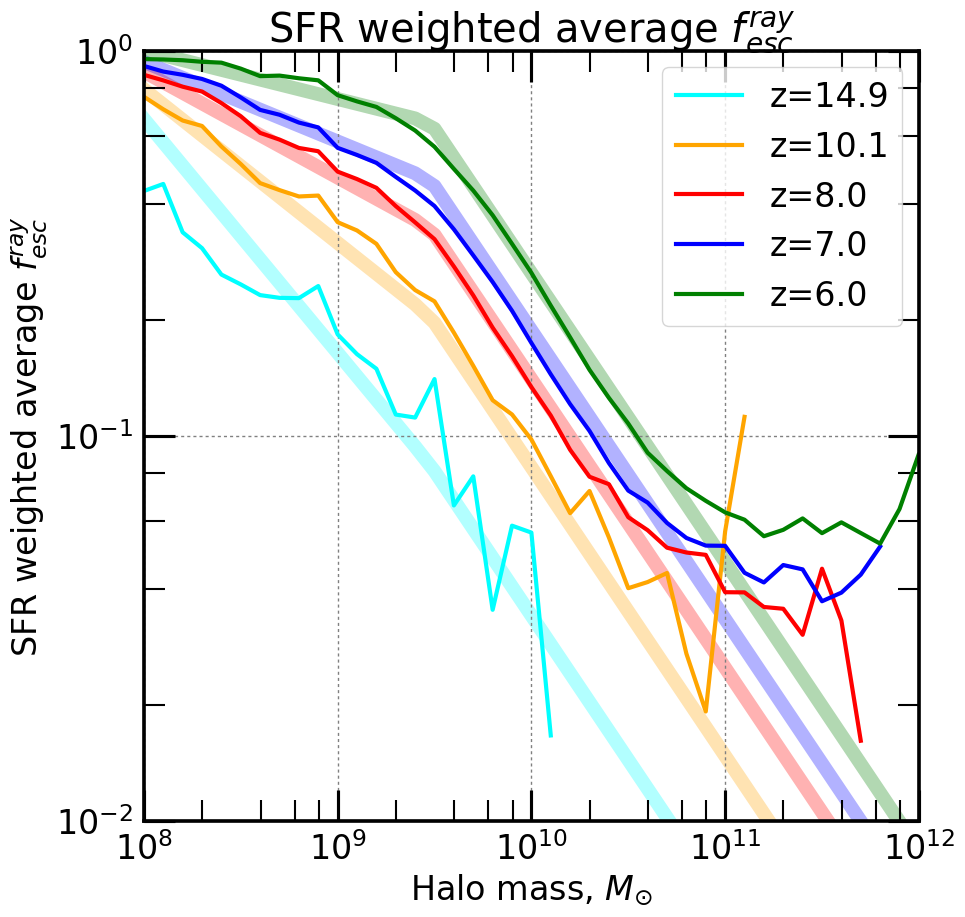}
   \caption{SFR-weighted average escape fractions (thin lines) of star-forming haloes and corresponding fits (thick lines) at z=6, 7, 8, 10.1, 14.9}
    \label{fig:fesc_SFR_fits}
  \end{figure}

We also provide fits for the SFR-weighted average values of $\fescr$ as a function of mass and of redshift between z=6 and z=14.9. As can be seen in Fig. \ref{fig:fesc_SFR_fits}, the SFR-weighted average of $\fescr$ has a slightly different behaviour. Indeed, in this case, the average curves do not join above a certain mass. To fit them, we opted for a slightly different double power law of halo mass, with a knee separating both laws. We propose the following functional form (Eq. \ref{eq:fesc_fct_SFR}):

\begin{equation}
    \fesc= \left \{
    \begin{array}[]{ll}
    \rm f^{\rm knee}_{\rm esc}(z)f^{\rm knee}(\mh,z); & \rm if \, \mh \geq M_{\rm knee}(\rm z) \, ,\\
    \rm f^{\rm max}_{\rm esc}(z)f^{\rm max}(\mh,z); & \rm if \, \mh < M_{\rm knee}(\rm z)\, ,
    \end{array}
    \right.
    \label{eq:fesc_fct_SFR}
\end{equation}

\begin{equation}
    \rm with= \left \{
    \begin{array}[]{lcl}
    \rm z_6 & = & \Big(\frac{1+z}{1+6}\Big)\, , \\
    \rm f^{\rm max}_{\rm esc}(z) & = & \rm min\big(\rm f^{\rm 0, max}_{\rm esc} z_6^{\beta},\,1.0 \big) \, , \\
    \rm f^{\rm knee}_{\rm esc}(z) & = & \rm min\big(\rm f^{\rm 0, knee}_{\rm esc} z_6^\gamma,\,1.0 \big) \, , \\
    \Delta(z) & = & \frac{\rm log_{10}(\rm f^{\rm knee}_{\rm esc}(z))-log_{10}(\rm f^{\rm max}_{\rm esc}(z))}{\rm log_{10}(M_{\rm knee}(z))-log_{10}(10^8)} \, , \\
    \rm M_{\rm knee}(\rm z) & = &3\times 10^9 \msol \, , \\
    \rm f^{\rm max}(\mh,z) & = & \Big(\frac{ \mh}{\rm M_{\rm knee}(\rm z)}\Big)^{\Delta(z)} \, , \\
    \rm f^{\rm knee}(\mh,z) & = & \Big(\frac{ \mh}{\rm M_{\rm knee}(\rm z)}\Big)^\delta \, , \\

    \end{array}
    \right.
    \label{eq:fesc_fct_SFR_params}
\end{equation}

$\rm f^{\rm max}_{\rm esc}(z)$ is the maximum value below $\rm M_{\rm knee}$, it is $f^{\rm 0, max}_{\rm esc}$ at z=6, and evolves as $z^\beta$.\\
$\rm f^{\rm knee}_{\rm esc}(z)$ is the where both power laws join at $\rm M_{\rm knee}(z)$, it is $f^{\rm 0, knee}_{\rm esc}$ at z=6, and evolves as $z^\gamma$.\\
$\Delta(z)$ is the slope of the power law below $\rm M_{\rm knee}(z)$, it is defined so as to reach $\rm f^{\rm max}_{\rm esc}(z)$ when $\mh=10^8\msol$.\\
$\rm M_{\rm knee}(\rm z)$ gives the mass where the power laws join, it is $3\times10^9 \msol$ at z=6, and evolves with z as $z^\zeta$.\\
$\rm f^{\rm max}(\mh,z)$ accounts for the mass slope for $\mh<\rm M_{\rm knee}(\rm z)$ given by $\Delta(z)$.\\
$\rm f^{\rm knee}(\mh,z)$ accounts for the mass slope for $\mh \geq \rm M_{\rm knee}(\rm z)$ given by $\delta$.\\

Since we want to reproduce the behaviour with mass and with redshift of the SFR weighted averages of $\fescr$, and not model the full distribution of points, we again adopt the simple approach of adjusting the fit by hand (as opposed to computing the fit of the model to the data sample).

The adjusted values are summarised in Eq. \ref{eq:fesc_fits_SFR_vals} below.
\begin{equation}
    \left\{
    \begin{array}[]{lrr}
    \rm f^{\rm 0, max}_{\rm esc} & = & 1.0 \, , \\
    \rm f^{\rm 0, knee}_{\rm esc} & = & 0.65 \, , \\
    \beta & = & -0.5 \, , \\
    \gamma & = & -2.5 \, , \\
    \delta & = & -0.75 \, , \\
  \end{array}
  \right.
  \label{eq:fesc_fits_SFR_vals}
\end{equation}

The aforementioned fits are presented in Fig. \ref{fig:fesc_SFR_fits} (full, thick lines), where they are also compared to the SFR-weighted average $\fescr$ measure in CoDa II (full, thin lines).

\section{Computing escape fraction}
\label{sec:fesc_draw}

\begin{figure}
\includegraphics[width=\columnwidth]{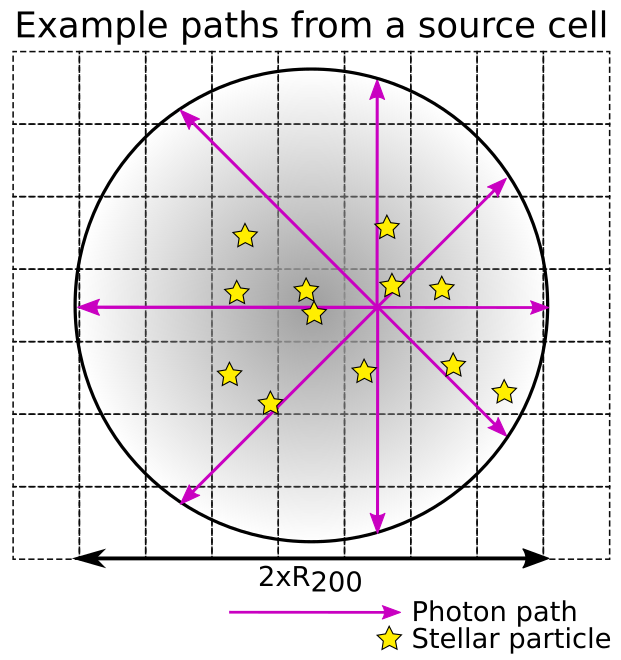}
   \caption{Explanatory drawing of the computation of $\fescr$. For each cell containing an emitting particle, we trace 768 rays from it's centre to a sphere of radius $\rtwo$ centred on the halo centre given by the fof halo finder. By averaging this result for every emitting cell and weighting by SFR, we obtain the halo value $\fescr$ at $\rtwo$.}
    \label{fig:fesc_draw}
  \end{figure}

Fig. \ref{fig:fesc_draw} shows a simplified, explanatory drawing of the computation process for $\fescr$ for an individual star forming halo cell.

\bibliographystyle{mnras}
\bibliography{new_photon_budget.bib} 

\bsp	
\label{lastpage}
\end{document}